\documentclass{aa}  

\usepackage{graphicx}
\usepackage{subcaption}

\usepackage{txfonts}
\usepackage{xcolor}
\usepackage[colorlinks=true, linkcolor=blue, citecolor=blue, urlcolor=blue]{hyperref}

\usepackage{natbib}
\usepackage[normalem]{ulem}

\usepackage{placeins}

\defcitealias{2021MNRAS.506.2269E}{EB21}

\begin{document}

   \title{Mind the companion: Demographics of transiting S-type exoplanets}

    \titlerunning{Mind the Companion}
    \authorrunning{L. Y. Messamah et al.}

    \author{Lina Y. Messamah, F. Bouchy, J. Venturini, L. Parc, A. Nigioni}

    \institute{University of Geneva, Department of Astronomy, Chemin Pegasi 51, 1290 Versoix, Switzerland\\
            \email{lina.messamah@unige.ch}
          }

    \date{Received Jan, 2026; accepted --- --, 2026}

    \abstract
    {Exoplanet demographic studies rely on large and homogeneous catalogues, yet stellar multiplicity remains incompletely characterised in many planet samples. Misidentified stellar companions can bias both stellar and planetary parameters, leading to ambiguous and incomplete conclusions about planet formation and evolution.}
    {We aim to construct a robust and reliable reference catalogue of S-type exoplanets, orbiting one component of a binary, for future investigations of planet formation and evolution in multiple-star environments, and to reassess exoplanet demographics by comparing planets hosted by single-stars and binary systems in a statistically consistent framework.} 
    {We updated the \textit{PlanetS} catalogue of transiting exoplanets by systematically identifying gravitationally bound stellar companions using Gaia Data Release 3 (DR3). Adopting a deliberately conservative classification, we distinguished binary and single-star systems and constructed a matched control sample of single hosts to mitigate selection and observational biases. Using this curated dataset of 860 transiting exoplanets including 133 S-type planets, we performed a comparative demographic analysis of planetary properties as a function of host multiplicity, stellar mass, and binary separation.} 
    {We find a binary fraction of 19.4\% relative to the control sample (15.5\% relative to the full single-star sample), consistent with previous estimates but derived from a larger and more homogeneous dataset. Significant demographic differences emerge in the giant planet regime, less affected by observational biases. We find that giant planets in binaries are more massive than their single-star counterparts and orbit closer to their host stars, making their radius more inflated. In particular, we derive that $\sim50\%$ of giant planets orbiting M-dwarfs are in binary systems, predominantly with separations $<1000$ AU, a $2.6\sigma$ excess compared to FGK-type hosts, suggesting that stellar multiplicity plays a key role in the formation or survival of giant planets around low-mass stars. In contrast, the small planet regime shows no statistically significant difference in binary fraction across stellar spectral types, although current statistics remain limited. We further explored correlations between planetary properties and host-star metallicity, finding trends consistent with core-accretion expectations largely driven by the giant planet population.}
    {}


   \maketitle

\section{Introduction}

Over the past three decades, the discovery of thousands of exoplanets has profoundly advanced our understanding of planetary system diversity and the mechanisms behind planet formation and evolution. The growing number of discovered planets highlights the need for thorough and complete databases that contain highly precise stellar and planetary parameters. 
These catalogues support population and demographic analyses, which aim to identify the statistical relationships between planet properties and stellar characteristics. Such studies are vital for validating and refining theoretical frameworks of planet formation across different stellar environments.

From a theoretical point of view, the presence of a stellar companion influences planet formation by truncating the protoplanetary disk \citep{1977MNRAS.181..441P,1994ApJ...421..651A,2019A&A...628A..95M,2021MNRAS.507.2531Z, 2026A&A...708A..38N, 2026A&A...708A..37V} and shortens its lifetime. This causes S-type planets\footnote{S-type planet: satellite planet, orbiting one component of the binary system} to be different than their counterparts around single stars. Observational evidence of this effect has been reported in the Kepler sample, where planets in tight binaries (separations < 100 AU) exhibit a suppressed sub-Neptune population and a unimodal radius distribution compared to planets around single stars or in wider binaries (separations > 300 AU) \citep{2023AJ....165..177S,2024AJ....168..129S}, suggesting that stellar multiplicity may fundamentally alter the planet formation process. The gravitational perturbations of the companion may also alter planetary migration or trigger dynamical instabilities at later stages \citep{2026A&A...708A..38N}.
On the other hand, very wide binaries (> 10 000 AU) are expected to have a negligible impact on the formation and evolution of planetary systems; nevertheless, they may still introduce photometric and spectroscopic contamination, which should be accounted for in analyses. 

A major source of uncertainty in current demographic studies arises from the presence of misidentified binary companions among planet-host stars. Many large catalogues, including the NASA Exoplanet Archive\footnote{\url{https://exoplanetarchive.ipac.caltech.edu}}, are constructed from heterogeneous sources that do not accurately flag binary companions or assert their physical associations. Recent compilations of planets in binary systems, such as that of \citet{2025A&A...700A.106T}, have provided valuable insights into the diversity of planetary architectures in multiple-star systems. However, these studies necessarily assemble targets from a variety of detection techniques and survey strategies, which makes it challenging to disentangle astrophysical trends from heterogeneous observational biases. A uniform and statistically robust distinction between single stars and binary systems, with reliable and consistent parameters remains lacking. As a result, the extent to which binarity influences the known population of exoplanets remains uncertain. 

The goal of this work is to establish a robust and reliable framework for distinguishing between single and binary hosts within the \textit{PlanetS} catalogue of transiting exoplanets \citep{2020A&A...634A..43O,2024A&A...688A..59P} and perform a demographic comparison between transiting exoplanets in single and binary systems.

Our goals are threefold: 
\begin{itemize}
    \item Identify and isolate gravitationally bound binary systems, based on robust and reliable astrometric parameters from Gaia data release 3 (DR3).
    \item Preserve robust and reliable planetary and stellar parameters, ensuring that the difference between binaries and single stars is not driven by measurement artefacts. 
    \item Enable a reliable statistical comparison between the two populations and investigate how planetary properties relate to binaries characteristics.
\end{itemize}

The \textit{PlanetS} catalogue (see Sect. \ref{sec. 2}), being the basis of this study, compiles well-characterised transiting exoplanets with reliable precisions on their radii and masses. Its robustness makes it suitable for such a differential study. Until now, the catalogue has not explicitly incorporated information about stellar companions, making this work the first attempt to quantify the impact of binarity on the demographics of transiting exoplanets within \textit{PlanetS}. We focus exclusively on transiting S-type planets in visual binaries identified in Gaia DR3; tighter unresolved systems are intentionally excluded and will be addressed in future studies.

This paper is organised as follows. In Sect. \ref{sec. 2}, we present the updated version of the \textit{PlanetS} catalogue, including the identification of stellar companions and the associated companion parameters for binary systems. Section \ref{sec:3} qualitatively describes the statistical sample and discusses the main observational and selection biases affecting it. In Sect. \ref{sec:4}, we introduce the methodology used to construct a matched control sample of single stars, enabling a robust comparison between planets hosted by single and binary stars by isolating the effect of stellar multiplicity from intrinsic stellar properties. The results of the comparative demographic analysis are also presented in this section.

\section{Binary identification and PlanetS catalogue update}
\label{sec. 2}
\textit{PlanetS}\footnote{ \url{https://dace.unige.ch/exoplanets/}} is a catalogue of transiting extra-solar planets with robust and precise mass and radius measurements (with mass precision $< 25 \%$ and radius precision $<8 \%$). It has been initially presented by \cite{2020A&A...634A..43O} and then extended and updated to include all masses alongside many other parameters -- including bulk density, insolation flux, and equilibrium temperature -- derived from homogeneous classical computations by \cite{2024A&A...688A..59P}.\\

The latest version of the \textit{PlanetS} catalogue (March 2026) counts 952 planets in total. In this work, we present an extension of the catalogue that explicitly accounts for transiting planets in gravitationally bound binary systems. Earlier versions of the catalogue, which relied on the NASA Exoplanet Archive, often contained misclassifications. Many systems were flagged as binaries without clear and robust evidence of being gravitationally bound, while others were listed as single stars and later found to have stellar companions. 
The goal of this update is therefore to provide a cleaner and homogeneous, although conservative, classification of planets in binaries, ensuring a solid foundation for future statistical and theoretical studies and enabling a more robust assessment of the impact of stellar multiplicity on planetary properties.
To classify the multiplicity of the stars, we cross-matched the \textit{PlanetS} catalogue with the catalogue of possible pairs in Gaia DR3 made by \citet{2021MNRAS.506.2269E} (hereafter \citetalias{2021MNRAS.506.2269E}). For each system, we ensured that the Gaia DR3 ID of the host would match either $source\textunderscore id1$ (the host is the primary component of the binary) or $source\textunderscore id2$ (the host is the secondary component of the binary system).
The \citetalias{2021MNRAS.506.2269E} catalogue was constructed from Gaia early data release 3 (eDR3 \citet{2021A&A...649A...1G}) data in order to identify 1.3 million spatially resolved binary stars with a high bound probability, within $\sim 1$ kpc of the Sun, with projected separations ranging from a few AU to 1 pc. \\
The candidate binaries were validated using near-identical parallaxes, consistent within 3 or 6 sigma, and proper motions consistent with a Keplerian orbit as described by the Eqs. 3-6 of \citetalias{2021MNRAS.506.2269E}. They also adopted a conservative approach by excluding clusters, triples, and candidate pairs in crowded regions. While this strategy improves the reliability of the resulting sample, it also means that a number of genuine binaries are inevitably excluded.\\

In the case of the \textit{PlanetS} catalogue, we selected binaries with projected separations $\lesssim10,000$ AU.
We adopted a strict limit on the chance-alignment probability ($R_{chance-align}$), as we selected only systems with a $R_{chance-align}< 0.01$ (i.e a physical association probability greater than 99\%). It should be noted that this cut is not overly restrictive, since there is a strong correlation between $R_{chance-align}$ and the projected separation. Essentially, all pairs with projected separations greater than $10,000$ AU have chance-alignment probabilities $>0.01$; therefore, no robustly bound wide binaries are excluded by this criterion. \\

For each system identified as a visual binary, we incorporated several key parameters to characterise the stellar companion. All the parameters as well as their description and sources are described in Table \ref{tab:binary_parameters}. 
We revised the column labelled, 'Number of stars', in the \textit{PlanetS} catalogue to reflect the updated multiplicity status of each system. This field was re-labelled 'Number of stellar companions' and redefined to indicate the number of gravitationally bound stellar companions identified through our approach (zero for single stars, one for binaries, and two for triples). This update accounts for systems previously classified as single and found to host a resolved stellar companion, as well as systems initially identified as multiple where no bound companion could be robustly confirmed after the crossmatch.

The angular separation ($\rho~[arcsec]$) of each pair was computed using the projected separation given by \citetalias{2021MNRAS.506.2269E} catalogue, and the parallax. The magnitude difference $\Delta G$ was included to identify the host star: $\Delta G > 0$ indicates the planet orbits the primary, while $\Delta G < 0$ indicates it orbits the secondary. \\

\begin{table*}
\footnotesize
\centering
\caption{Parameters included for each system identified as a visual binary in the updated version of the PlanetS catalogue.}
\begin{tabular}{llp{5cm}llp{2.5cm}}
\hline
\textbf{Parameter} & \textbf{Unit} & \textbf{Description} & \textbf{Source} & \textbf{Value range} &  \textbf{Number of available values} \\
\hline
Number of companions & -- & Number of stellar companions identified for the host & \textit{This work} & [0;1] & 133  \\
Angular separation & arcsec & Angular separation between host and companion & \textit{This work} &  [0.71;68.9] & 133 \\
Projected separation & AU & Physical projected separation of the binary  & \citetalias{2021MNRAS.506.2269E} &[76;9655] & 133 \\
$G_{\mathrm{companion}}$ & -- & Gaia G-band magnitude of the stellar companion & \textit{Gaia DR3} & [8.9; 20.8] & 133\\
$B_p - R_p$ & -- &  Gaia colour index of the companion & \textit{Gaia DR3} & [0.4; 4.0] & 120 \\
$\Delta G = G_{\mathrm{companion}} - G_{\mathrm{host}}$ & -- &  Gaia magnitude difference between the companion and host. & \textit{Gaia DR3} & [-4.7; 11.6] & 133 \\
Companion $T_{\mathrm{eff}}$ & K &  Effective temperature of the companion & \textit{Gaia DR3} & [2083; 6087] & 133 \\
Companion $\mathrm{[Fe/H]}$ & dex & Metallicity of the companion &\textit{Gaia DR3} & [-1.25; 0.75] & 59\\
$L_{comp}$ & $log(L_{\odot})$ & Luminosity of the companion & \textit{Gaia DR3} & [-2.73;0.44] & 52 \\
$\log g_{comp}$ & $log(cm/s^2)$ & Surface gravity of the companion & \textit{Gaia DR3} & [4.9 ; 5.1] & 59\\
\hline
\end{tabular}
\label{tab:binary_parameters}
\end{table*}

We emphasise that the classification of a system as 'single' relies on the criteria used in this work and does not exclude the potential presence of undetected or unconfirmed stellar companions. In particular, several targets identified as single do exhibit candidate companions in direct imaging data or in sparse ground-based astrometric measurements. However, in these cases, the available information is not conclusive enough (according to our adopted criteria) to confirm a gravitationally bound association with the host star. For directly imaged companions, the lack of parallax and proper motion measurements for the companion star prevents us from reliably confirming common movement. As for astrometric companions detected from the ground, the time baseline and low number of data points are not sufficient to rule out chance alignment with statistical confidence.
Future data releases, such as Gaia DR4/DR5, may allow us to re-evaluate such cases with improved astrometric precision and longer baselines. As of now, our classification adopts a conservative approach, ensuring that only well-characterised systems are included in the binary sample.\\

The original version of the \textit{PlanetS} catalogue included 15 triple systems. As part of this work, we revisited these cases individually to ensure a consistent and astrophysically reliable classification with the Gaia DR3 framework. Based on this reassessment, we updated the classification of several systems in the catalogue; the final adopted number of companions are summarised in Table \ref{tab:triple_systems}. 
Among these, only one triple system remains classified as a confirmed triple system in our catalogue: LTT-1445, which hosts two transiting planets initially validated and confirmed by \cite{2019AJ....158..152W} and \cite{2022AJ....163..168W}. Later on, \cite{2023A&A...673A..69L} refined the mass of these planets. \cite{2022AJ....163..168W} mention that it is a triple system of M-dwarfs, where the unresolved BC pair induces an astrometric perturbation on the primary component, which was revealed through 18 years of ground-based astrometric monitoring. \cite{2024AJ....167...89Z} also provide a detailed characterisation of the 3D orbital architecture using radial velocities and absolute astrometry from Gaia and Hipparcos, finding a nearly coplanar configuration with a mutual inclination of $\sim~2.88\deg$ between the orbit of the BC pair around A and that of C around B. Gaia DR3 does not provide astrometric solutions for the BC pair. For this reason, there is no additional information included on this system in the \textit{PlanetS} catalogue.  A detailed justification for the adopted classification in our work for the remaining 14 systems, including the criteria used to discard or revise their triple status, is provided in Appendix \ref{Appendix A}.

\section{Sample properties and statistics}
\label{sec:3}
\subsection{Description of the sample}

The results presented hereafter are based on the version of the \textit{PlanetS} catalogue as of August 2025 including a total of 860 planets, which constitutes the version on which the statistical analysis was performed. Following the methodology and applying the criteria described in Sect. \ref{sec. 2}, we identify 133 transiting planets residing in gravitationally bound binary systems. The remaining sample consists of 725 planets hosted by apparently single stars, along with two planets belonging to the hierarchical triple system LTT 1445.

Among the 133 planets identified in binaries, only five of them are hosted by the fainter component in $G_{mag}$. In addition, nine systems have $|\Delta G| < 1$, indicating that they are near-equal-magnitude binaries, assuming that they are both main-sequence stars, and at the same distance, we could infer that they are near-equal-mass binaries.\footnote{For FGKM dwarfs, a $|\Delta G|$ of 1 corresponds at maximum to a $25\%$ difference in mass.} A complete summary of the distributions of all parameters is provided in Table \ref{tab:binary_parameters}. Note that the number of available values denote the number of planets in the \textit{PlanetS} catalogue for which those parameters are available, and not the number of planetary systems; therefore, multi-planetary systems may be counted more than once. \\

\begin{figure}
\centering
\includegraphics[width=\linewidth]{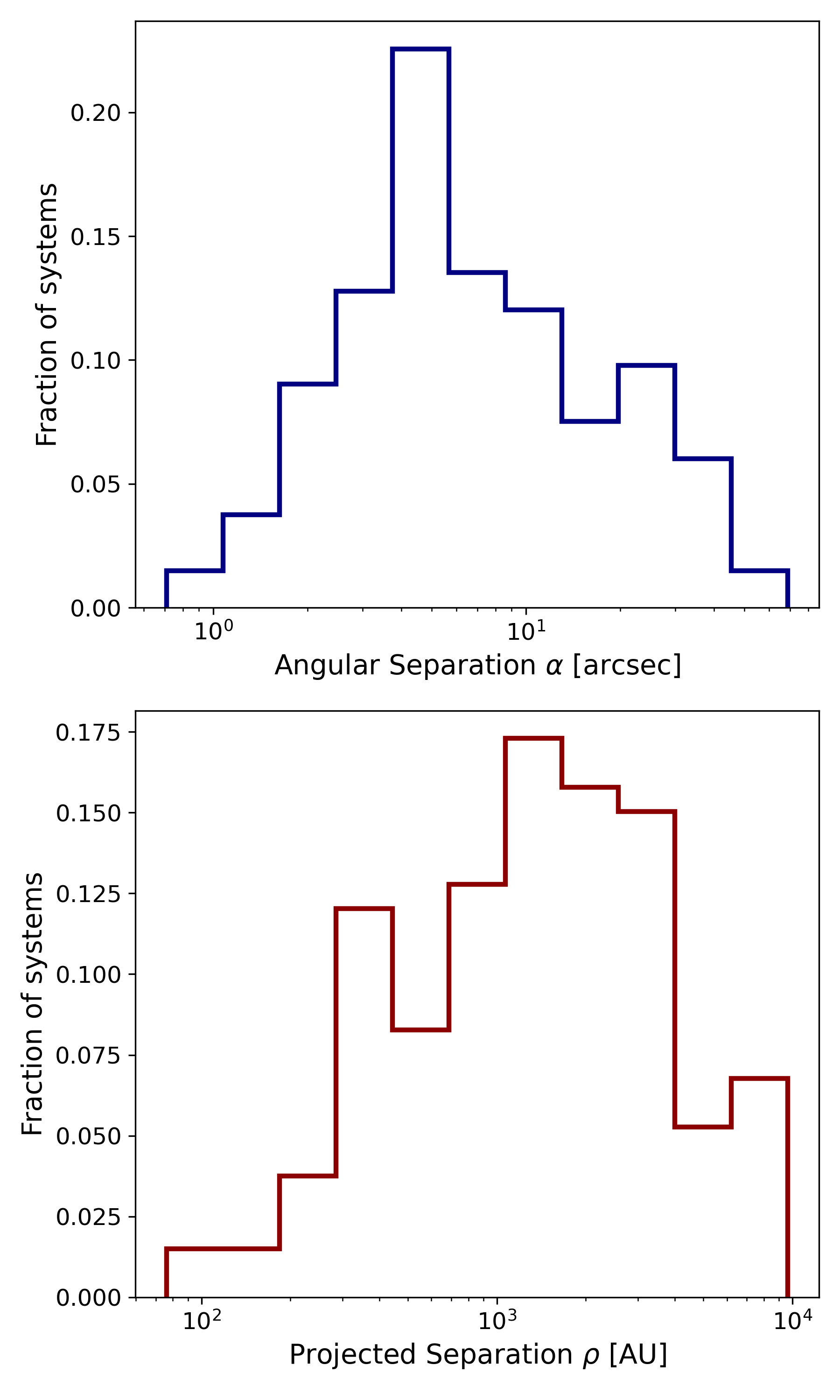}
    \caption{\small Distribution of angular separations (top panel) and projected separations (bottom panel) of the binary sample in the \textit{PlanetS} catalogue.
          }
        \label{fig: separations}
\end{figure}

To better understand the properties of the binary sample, we examined the distributions of angular separations and projected separations, as shown in Fig. \ref{fig: separations}. The angular separations distribution has a median value $\sim 5.7$ arcsec, while the projected separations have a median value around 1400 AU. The binary parameter space probed in this study spans projected separations from 76 AU to 10'000 AU, with the lower limit driven by Gaia's angular resolution and the upper limit imposed by our chance alignment criterion ($R_{chance-align} \leq 0.01$). Observational biases are discussed in more details in the following section.

There is no unique convention in the literature for the boundary between ‘close’ and ‘wide’ binaries; thresholds from tens to hundreds of AU are commonly used, motivated by different physical considerations and datasets. In the following, we refer to visual binaries with projected separation s < 1400 AU as ‘close’ and those with s > 1400 AU as ‘wide’; this boundary corresponds to the median projected separation of our sample and was adopted for convenience between two comparable subsamples. Furthermore, this boundary corresponds to the upper limit of binaries separation impacting planet formation  \cite{2014ApJ...791..111W}.

\subsection{Observational and selection biases}
\label{Sec. 3.2}
Our sample consists of transiting exoplanets with masses derived from radial velocity (RV) follow-up or timing transit variations (TTVs) for some cases. For the latter, only planets with robust mass estimations against mass and eccentricity degeneracy were considered \citep{2024A&A...688A..59P}.
In the case of binary systems, the presence of a nearby stellar companion can introduce observational biases. 
On the photometric side, additional flux from a contaminant dilutes the transit depth, leading to underestimated planetary radii, as shown in Eq. \ref{eq:dilution factor} (see \citet{2015ApJ...805...16C} for the derivation):

\begin{equation}
    \delta_{obs} = \frac{\Delta F}{F_{total}} = \bigg(\frac{R_p}{R_{*}}\bigg)^2 . \frac{F_{host}}{F_{host}+F_{comp}}.
    \label{eq:dilution factor}
\end{equation}
This effect is strongest when the flux ratio between the companion and host is high, and the radius is underestimated. So, the true radius of the planet would be
\begin{equation}
    R_{p,true} = R_{p,obs}.\sqrt{1+\frac{F_{comp}}{F_{host}}}
    \label{eq: true radius}.
\end{equation}

Moreover, neglecting flux contamination can significantly bias planetary radius measurements. \citet{2025ApJ...988L...4H} show that blending with background or nearby stars, when not properly corrected, leads to systematically underestimated planetary radii. Their results indicate that radii derived from uncorrected TESS photometry are, on average, underestimated by about 6\%. This bias affects all TESS targets in a similar way, whether the host star is single or in a binary system.

On the spectroscopic side, light from the stellar companion may contaminate the spectra injected in the spectrograph fibre, biasing the inferred RV signal, which in turn affects the measured planetary mass. These effects should be accounted for when assessing the demographics of exoplanets in binaries, as they can significantly distort the observed distributions. This is especially true for small, low-mass planets, which are more sensitive to small errors, potentially rendering them undetectable or leading to systematic underestimation of their properties.

For all the published transiting planets with a reliable mass and radius measurement, part of the PlanetS catalogue, all the contaminants, and field stars are known from previous astrometric surveys and/or previous Gaia releases and are taken into account in the photometric extraction and dilution factor correction. We verified that for the five systems in our binary sample with angular separations < 2 arcsec published before Gaia DR3, the stellar companion was already known and accounted for in the derivation of the planetary parameters. These contaminants, as well as the faintest ones revealed by Gaia DR3, have no impact on radius estimation. Our present work determines whether these close contaminants are gravitationally bound stellar companions. On the other hand, our selection introduces an inherent bias towards short-period planets, as our planets are characterised with the transit and RV methods.

\begin{figure}[htbp]
\centering
\includegraphics[width=9.5cm]{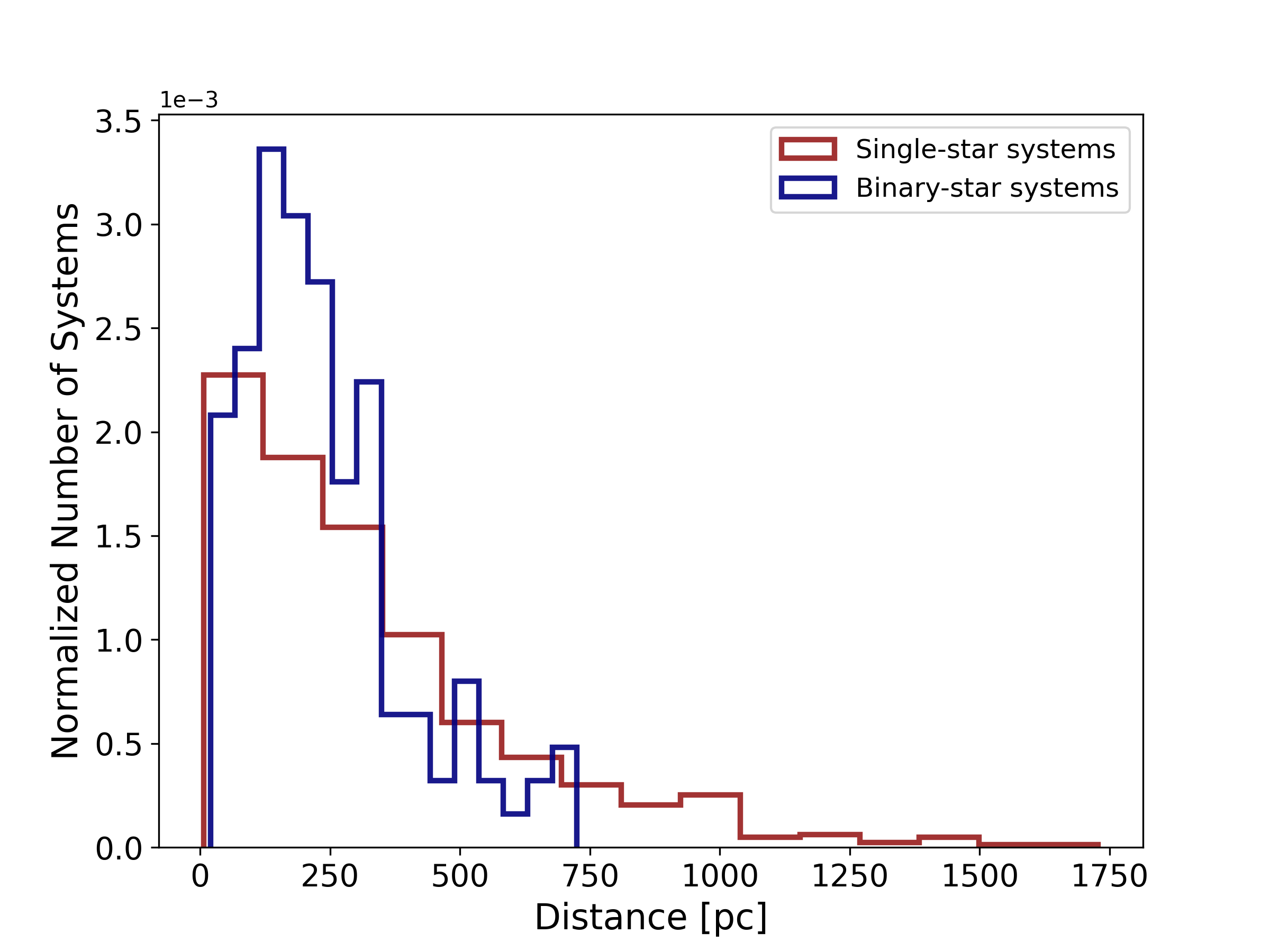}
    \caption{\small Distribution of distances to the sun in parsec for binaries (orange) and single stars (blue).
          }
        \label{Fig:distances}
\end{figure}

To assess the resulting selection bias, we compared the distance distributions of planet-hosting binary systems and single-star systems (Fig. \ref{Fig:distances}). We find that binaries span a slightly narrower range of distances, while single-star systems peak at closer distances to the Sun, extending farther out.
A Kolmogorov-Smirnov (KS) test confirms that the two samples differ significantly ($D=0.194$, $p=5.8 \times 10^{-4}$). This difference reflects the angular resolution and contrast sensitivity of Gaia DR3, illustrated as a function of angular separation and magnitude contrast by (\citet{2024MNRAS.527.3183M}, and \citet{2021FrASS...8...14M}), where Gaia's detection sensitivity drops sharply below $\sim0.7$ arcsec. This means that companions at projected separations below $\sim 70 $ AU at typical distances in our sample remain undetectable. It moreover implies that our sample is incomplete for tight and/or faint stellar companions and that the observed deficit of systems at angular separations $< 1-2$ arcsec (top panel of Fig. \ref{fig: separations}) reflects this detection boundary rather than a physical absence of such configurations. As a result, resolved binaries are preferentially detected at intermediate distances, while apparently single systems populate both very near and very distant ends of the distribution.
One indicator of potentially unresolved binarity provided in Gaia DR3 is the renormalised unit weight error (RUWE), which quantifies the quality of the single-star astrometric fit. Sources with high RUWE values are considered to have poorly constrained astrometric solutions, potentially indicating the presence of an unresolved companion. We examined the RUWE distributions of the sources in our sample and found only a small fraction exhibit significantly elevated values. A detailed discussion of the RUWE distributions and their implications for our sample is provided in Appendix \ref{Appendix ruwe}.

\section{Demographic comparison: Single versus binary hosts}
\label{sec:4}

\subsection{Control sample and global comparison}
\label{Sec 4.1}
To compare the population of exoplanets in binaries to that around single stars accurately, we aimed to build a control sample of single stars as a subsample of the full population of single host stars. The purpose of this control sample is to remove potential biases introduced by differences in stellar properties, ensuring that any observed differences between the two populations can be attributed to binarity rather than to underlying stellar characteristics. 

To achieve this, the control sample was designed to reproduce the distributions of key stellar parameters observed in the binary-host sample. We matched the distributions of stellar effective temperature, metallicity, stellar mass, and distances. This was done through a 4D bubble-matching approach in the stellar parameter space, which consists of the following steps. First, the binary hosts and single-star host populations was first mapped in this 4D parameter space to define the reference distributions. Each feature was then standardised using the \texttt{StandardScaler} routine from the \texttt{sklearn} python library \citet{scikit-learn}, ensuring that each parameter contributed equally in dimensionless units. The scaler was fitted on the binary hosts population to preserve its statistical structure and then applied consistently to the single hosts sample. 

\begin{figure}
\centering
\includegraphics[width=9cm]{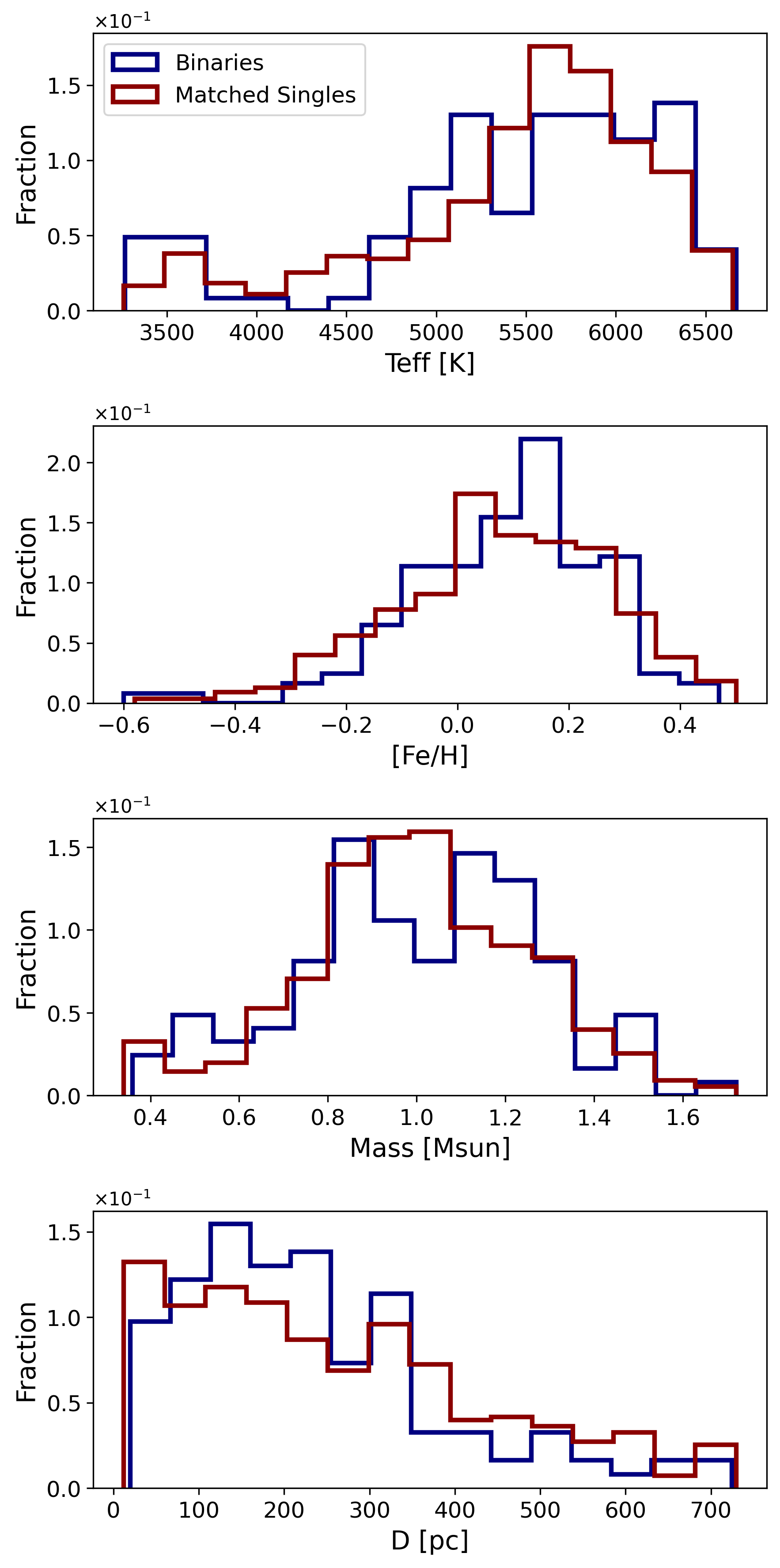}
    \caption{\small Stellar effective temperature, metallicity, mass, and distance distribution for the stellar binary and control samples.
          }
        \label{fig:control_sample_distribution}
\end{figure}

We then computed the centroid of the binary population in the standardised parameter space and used it as the reference centre for our 4D bubble. Instead of defining a single spherical radius, we adopted custom tolerances in each direction, given the difference in distribution of the stellar parameters. The distances were computed using the \texttt{cdist} function of \texttt{Scipy} \citep{2020SciPy-NMeth}. Single stars whose standardised parameters fell into those boundaries were retained as our matching control sample. Finally, we verified (see Fig. \ref{fig:control_sample_distribution}) the matching quality by comparing the resulting normalised distributions of the four parameters for both samples.
The resulting distributions show consistent medians and variances, confirming that the selected control sample presents a statistically representative baseline for demographic comparison.

We obtain a control sample of 552 planets in single-star systems, and comparing it to the binaries sample, we estimate the overall fraction of planets hosted in resolved binary systems to be ~$19.4\%$, computed as $r_B = (N_B/N_{tot})$, where $N_B$ is the number of planets in binaries, and $N_{tot}$ the total number of planets in the sample (binaries plus control single systems), later referred to as 'binary ratio' or 'binary fraction'. Although this global value indicates a relatively low incidence of binarity in our sample, it should not be interpreted as evidence that planets are intrinsically less common in binary systems. Our sample is inherently biased by observational selection effects, as discussed in Sect. \ref{Sec. 3.2}. Instead, we treated the global binary fraction as a baseline expectation, representing the overall prevalence of binarity within our observed sample. 

In the following, we compare key planetary properties between the binary and single-host populations to explore whether binarity correlates with differences in planet mass, radius, orbital period, and other characteristics. From now on, our full analysis is performed by making comparisons with the control sample of single stars, unless specified otherwise. \\

\begin{figure}
\centering
\includegraphics[width=9.5cm]{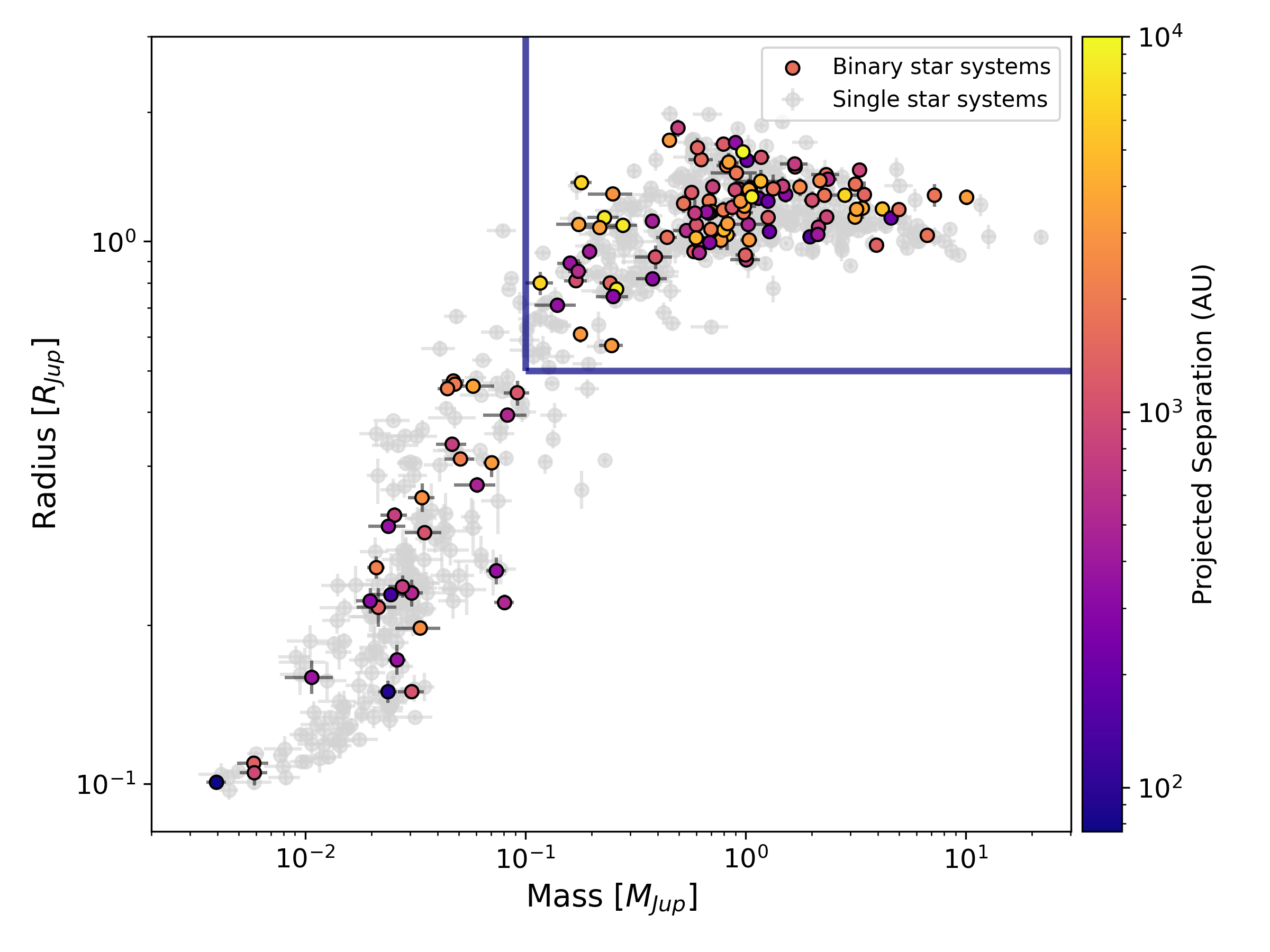}
    \caption{\small M--R diagram of the transiting exoplanet population in binary systems, colour-coded by projected separation (AU), and the control sample of transiting exoplanet in single-star systems (grey). The blue rectangle delimits the regime of giant planets (radius $\geq0.6$ $R_{Jup}$ and mass $\geq0.1$ $M_{Jup}$), distinguishing them from the rest of the populations.
          }
        \label{MRfig}
\end{figure}

We begin this global comparison by examining the distribution of planets in the mass--radius (M--R) diagram, shown in Fig. \ref{MRfig}. This diagram shows that binary-hosted planets span the full range of planetary masses and radii, from rocky small planets to more massive gas giants. We notice that the M--R envelope is similar between populations of planets around single stars and in binary systems. No significant segregation is immediately apparent based on binary separation as planets in both close, and wide binaries appear across all planet regimes. 

\begin{figure*}[htbp]
\sidecaption
    \includegraphics[width=12cm]{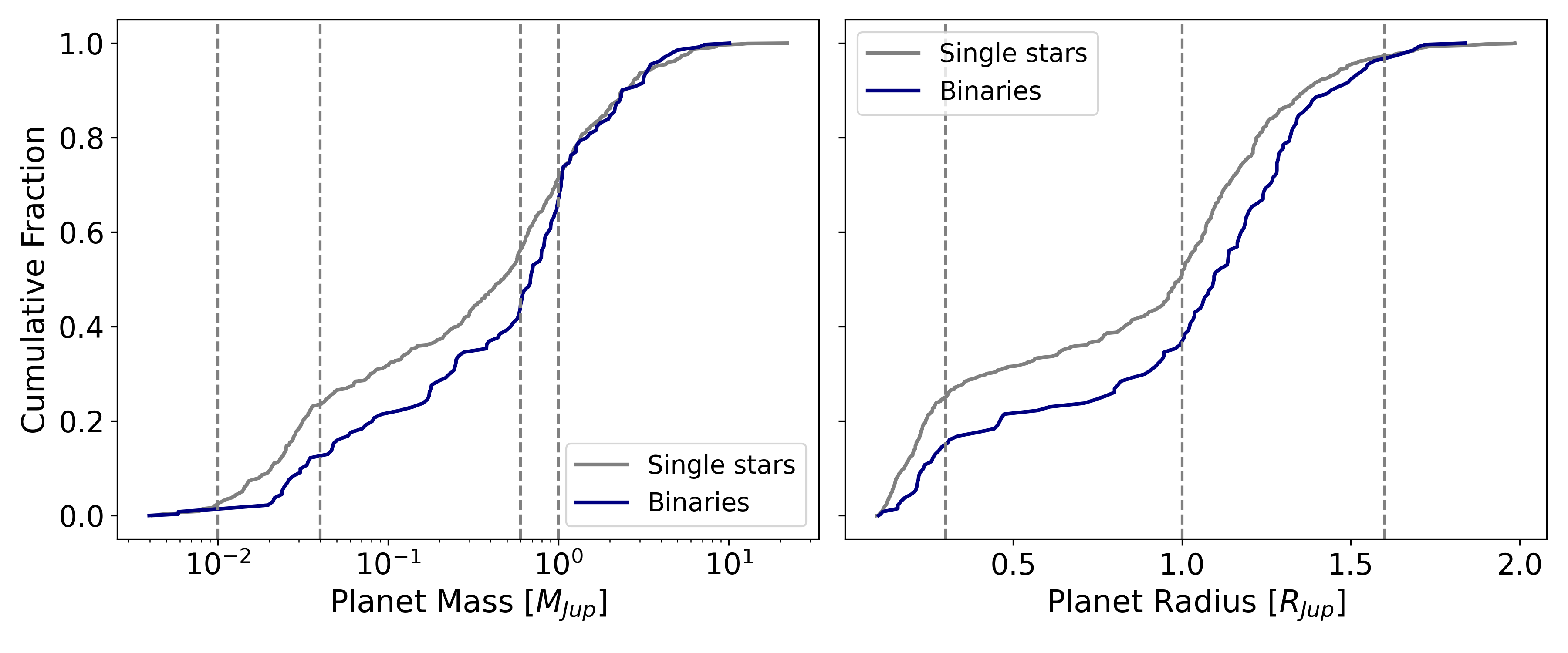}
    \caption{\small Cumulative distributions of planetary properties in binary systems (blue) and single stars (grey), with the planet mass distributions (left) given in Jupiter masses and the planet radius distributions (right) given in Jupiter radii.
    The dashed lines correspond to regions where the main differences in the distributions arise, as discussed in Sect. \ref{Sec 4.1} }
    \label{fig:CDFs_mass_radius_all}
\end{figure*}

To statistically assess whether planets in binary systems differ from those orbiting single stars, we compared the cumulative distributions of planet radius and planet mass between the two populations. Figure \ref{fig:CDFs_mass_radius_all} represents these cumulative distributions for planet radii (right) and planet masses (left). The significance of the difference between the two samples is assessed using a Mann-Whitney U test. For planet masses, the test yields a p-value $0.027$, suggesting a relatively significant difference between the two distributions. The median mass of the planets orbiting single stars is $0.47\pm0.06~M_{Jup}$, while the median mass for binary-hosted planets is $0.69\pm0.08~M_{Jup}$, a difference significant at the $2.2\sigma$ level, which indicates that planets hosted in binary systems are slightly skewed towards higher masses compared to single-star systems.

Consistent with the findings of \citet{2024MNRAS.527.3183M}, this divergence arises mainly in two regions of the distribution: at low masses from $0.01$ to $0.04$ $M_{Jup}$ ($3.18 - 12.7 ~M_{\oplus}$), where single-star planets dominate. This is likely a consequence of observational and selection biases affecting the mass measurement for planets in close binaries ( angular separation typically $\lesssim~ 2 ~arcsec$), as mentioned in Sect. \ref{Sec. 3.2}. In contrast, at intermediate to higher masses ($0.6$ - $1 $$M_{Jup}$) binary-hosted planets are comparatively more frequent, suggesting a genuine difference in the underlying planet populations of that regime.

A similar tendency was already reported by \citet{2012A&A...542A..92R}, who found that their sample of planets -- detected, similar to ours, via transits and radial velocities -- also showed that binary-hosted planets are systematically more massive than those around single stars. Our results show that even with a substantially larger sample, we were able to recover the same trend, suggesting that binaries may indeed preferentially host more massive planets. However, our sample remains dominated by giant planets, while small-planet regime is strongly affected by detection biases. Thus, definitive conclusions regarding low-mass planets cannot yet be drawn, even though the giant planet population provides a more robust basis for exploring the influence of binarity on planet formation.  \\

In the case of the distribution of planet radii as shown in the right panel of Fig. \ref{fig:CDFs_mass_radius_all}, we conduct a similar analysis to the one described above. The Mann-Whitney U test yields a p-value of 0.0004, rejecting the null hypothesis and suggesting a statistical difference between the two populations. This time, the median radius of planets in binary systems is $\widetilde{R}_b = 1.095\pm0.035~R_{Jup}$, while planets around single stars have a smaller median radius of $\widetilde{R}_s = 0.992\pm0.017~R_{Jup}$, a difference significant at the $2.6\sigma$ level, which can indicate that planets in binary systems tend to have slightly larger radii. We can distinguish two regions in which the difference between the two distributions mainly arises. For small radii $\lesssim 0.3 R_{Jup}$ ($3.36~R_{\oplus}$), we notice that single star systems are more prevalent; however, this cannot be indication enough that small planets cannot be formed in binary systems. This difference is likely dominated by observational biases due to dilution effects in the light curves (see Sect. \ref{Sec. 3.2}). For larger radii $1-1.6~ R_{Jup}$, we notice that binary systems are more prevalent in this region. Given that our sample is inherently biased towards the detection of larger planets, this excess cannot be attributed solely to observational incompleteness and likely reflects a genuine demographic difference.

\subsection{Giant planets }
\label{sec:4.2}
As demonstrated in the previous section, the regime of small to intermediate planets, as defined in Fig. \ref{MRfig}, is heavily affected by observational and selection biases. As a result, binaries are significantly underrepresented in this part of the parameter space.
To ensure a more reliable and less biased comparison between binary and single star systems, we now focus our analysis on the region highlighted by a blue square in Fig. \ref{MRfig}, which corresponds to a range of planet mass and radius where detection and selection biases are minimal. This region is delimited based on the trends and biases identified in the sections above and include planets with masses $M_p\geq0.1$$M_{Jup}$, adapted as a lower mass limit for Jovian planets (\citet{2018haex.bookE.143M}; \citet{2021FrASS...8...16F}) as well as radii $R_p\geq0.6$$R_{Jup}$. This region is well-populated with 102 giant planets in binaries and a binary-to-single ratio of $22.06 \pm 0.14\%$ of the planets in that region. This makes it a suitable domain for a more focused and statistically meaningful comparison of planet demographics between the two populations. \\

In this regime, we find a median planet radius of $\widetilde{R}_b = 1.187\pm0.029~R_{Jup}$ for binaries versus $\widetilde{R}_s=1.115\pm0.014~R_{Jup}$ for single-star systems ($2.3\sigma$ significance level and a Mann-Whitney p-value $=0.024$). By contrast, the median planet mass shows no statistically significant difference (agreeing within $0.2\sigma$), suggesting that the mass difference observed in the full sample (Sect. \ref{Sec 4.1}) is not driven by  atendency of giant planets being intrinsically more massive. Rather, it reflects the higher proportion of giant planets relative to small planets in the binary sample compared to the single-star sample. 

\begin{figure*}
\sidecaption
    \includegraphics[width=12cm]{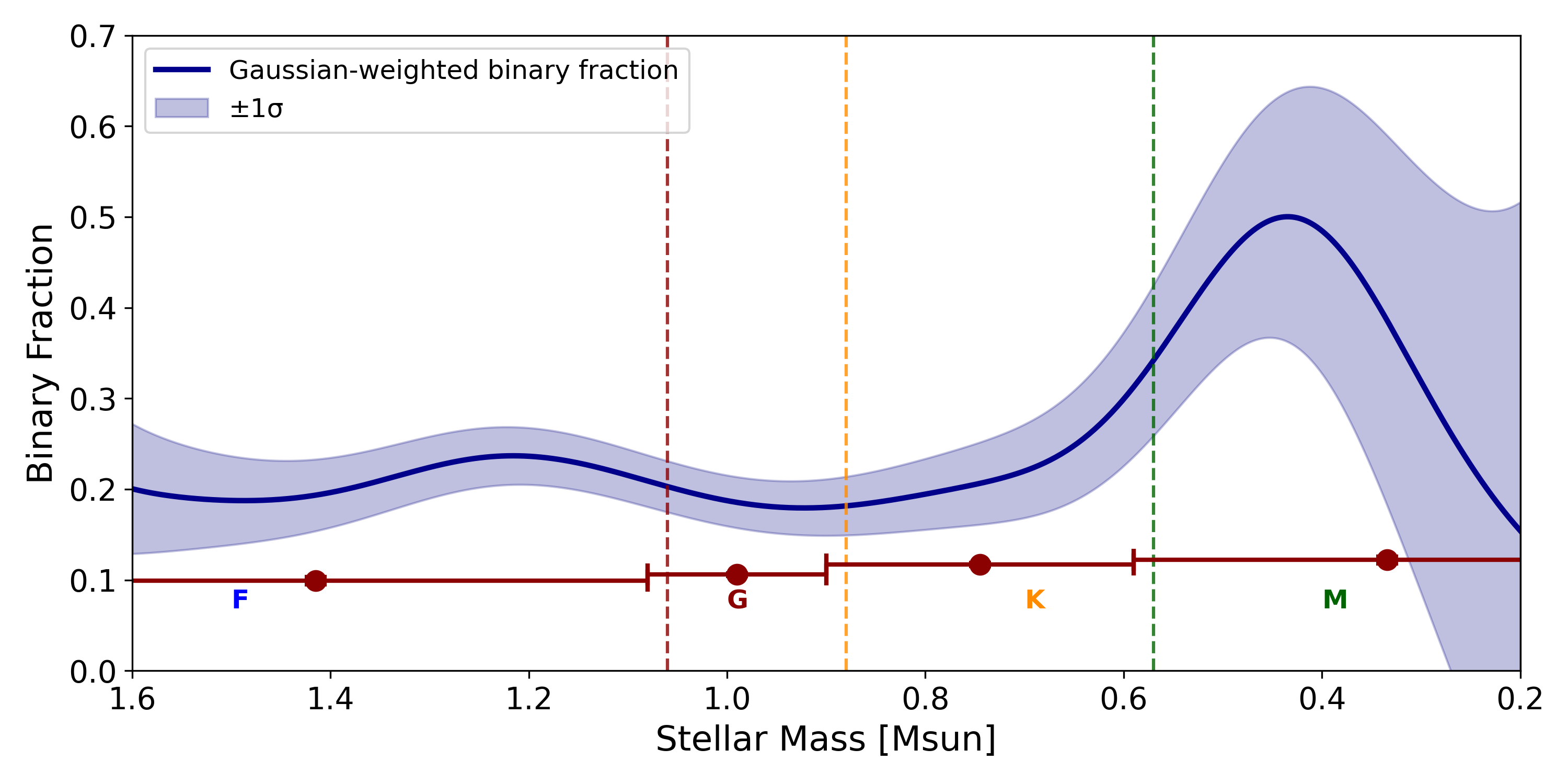}
    \caption{\small Fraction of giant planets in binary systems as a function of stellar mass [$M_{\odot}$] (blue line). The shaded region shows the 1$\sigma$ binomial uncertainty on the weighted fraction. The vertical dashed lines mark boundaries between the F, G, K, and M spectral types, as given in \citet{2013ApJS..208....9P}. The red dots represent the known multiplicity fraction of field stars as computed from Gaia DR3.}
    \label{fig:bin_frac_giants}
\end{figure*}

We aim at investigating the effect of the companion on the formation of giant planets through spectral types. Only one giant planet is catalogued around an A-type star ($T_{eff} = 7800~K$) corresponding to the MASCARA-4 binary system. This spectral type range is therefore not included in our analysis, as it will most likely skew our statistics. Figure \ref{fig:bin_frac_giants} shows the variation of binary fraction with respect to stellar masses. To achieve this, we adopted a Gaussian kernel of width $\sigma=0.1~M_{\odot}$, evaluated at 500 equally spaced steps across the full range of stellar masses. At each grid point $M_0$, each planet host $i$ contributes a Gaussian weight:

\begin{equation}
    \omega_i =exp\left(-\frac{(M_{*,i}-M_0)^2}{2\sigma_{eff,i}^2}\right),
\end{equation}
where the effective width $\sigma_{eff,i}=\sqrt{\sigma^2+\sigma^2_{M_{*,i}}}$ incorporates both the smoothing kernel and the individual stellar mass uncertainty. The fraction in each bin is computed as the number of planets in binaries with respect to the total number of planets (binaries plus singles). The shaded region corresponds to the $1\sigma$ uncertainty, on the binaries fraction following binomial statistics (\citet{2017ApJS..230...15M}, \citet{2023ASPC..534..275O}) and including the error on the estimated stellar mass, with the boundaries between spectral types extracted from \citet{2013ApJS..208....9P}. We note that this curve represents a lower limit to the binary fraction in our sample, as many apparent single-star systems could be unresolved binary systems that have not been accounted for as such. 
The red points represent the binary fraction among Gaia DR3 field stars, computed using the same approach applied to the planet host sample. Specifically, we selected main-sequence AFGKM stars from Gaia DR3 and applied a 2D bubble-matching approach in the space of distance and G-band magnitude, analogous to those used to construct the control sample (Sect. \ref{Sec 4.1}), to ensure that the field star sample shares the same observational characteristics as our planet hosts. We identified 13,148 gravitationally bound companions using the \citetalias{2021MNRAS.506.2269E} catalogue with identical criteria than explained in Sect. \ref{sec. 2} among 118,650. We find an overall binary fraction of $\sim11~\%$, slightly lower than the fraction recovered for planet hosts, broadly consistent across spectral types.
For comparison, the literature reports significantly higher overall multiplicity fractions, as reported in multiple papers described in the Figure 1 and Table 1 of \citet{2023ASPC..534..275O}, showing that the binary fraction decreases when decreasing the stellar host's mass. \citet{2010ApJS..190....1R} find a multiplicity fraction of $\sim~44\%$ for stellar masses between $0.75 - 1.25 ~M_{\odot}$ within 25 pc, while \citet{2019AJ....157..216W} reports $\sim~26\%$ for M-dwarfs in the same volume. These values are higher than what we recover in this work. This is expected given that the \citepalias{2021MNRAS.506.2269E} catalogue is sensitive to visual binaries with large projected separations and is therefore incomplete with respect to shorter period binaries. On the other hand, while the literature consistently reports a decreasing multiplicity fraction with decreasing stellar mass \citep{2010ApJS..190....1R,2019AJ....157..216W,2021MNRAS.507.3593M}, our binary field stars sample shows a relatively flat distribution with spectral type.

The fraction of planets binaries in our sample for F, G, and K-type is essentially constant within uncertainties. 
Our results also indicate that $\sim 50\%$ of giant planets orbiting M-dwarfs are found to be in binary systems. This represents a significant excess compared to other spectral types, where the binary fraction remains well below 30\%, as well as a significantly larger fraction than field binaries for low-mass stars, which we compute to be $12.2\pm0.4~\%$. This trend, also reported in \citet{2026A&A...707A..73F}, likely suggests a spectral-type dependence in how binary companions affect giant planet formation or survival. In addition, we find that a majority ($55.6~\%$) of the M-dwarfs stellar companions lie within 1000 AU, compared to only $30-40~\%$ for higher-mass host stars. 

Low-mass stars are known to host fewer giant planets with an occurrence rate between 0.05\% and 0.3 \% depending on sample and period range \citep{2022A&A...664A.180S,2023MNRAS.521.3663B,2026AJ....171..146G}. This supports the idea that M-dwarfs possess less massive protoplanetary disks \citep{Andrewsetal.2013,Almendros-Abadetal.2024}, limiting the efficiency of core accretion and subsequent gas accretion. However, our results show that although those systems are rare, when they exist, they tend to be hosted in close binary systems. 
This might suggest dynamical interactions induced by the stellar companion playing a role on the formation or evolution of the systems, but the reason is still unknown and deserves further investigation.\\
\begin{figure}
    \centering
    \includegraphics[width=9cm]{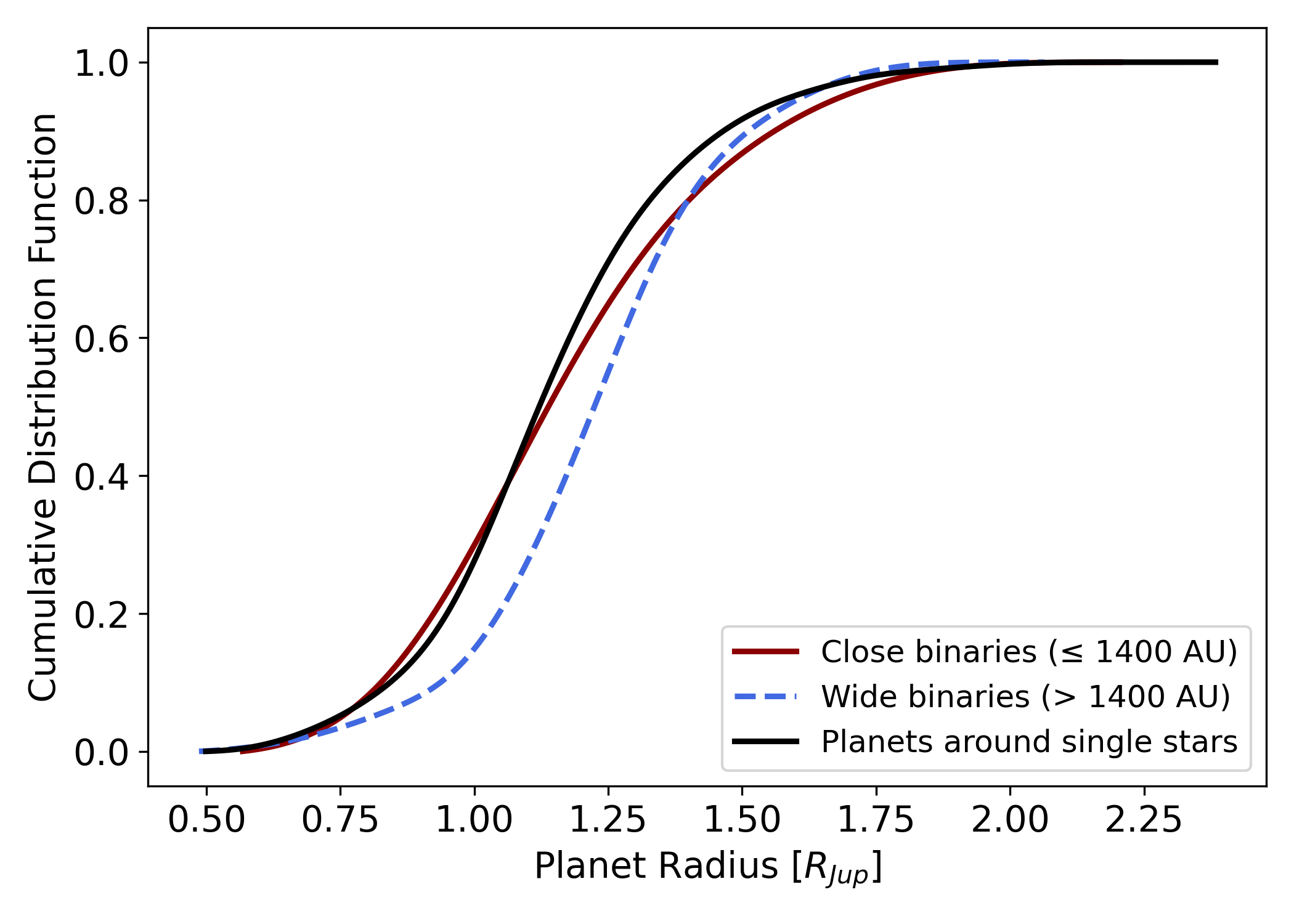}
    \caption{Cumulative distributions giant planet radii for close binaries (< 1400 AU) (red), wide binaries (blue), and single-star systems (black).}
    \label{fig:CDF_radii_giants}
\end{figure}

When comparing the cumulative distributions of radius for giant planets as shown in Fig. \ref{fig:CDF_radii_giants}. We find that planets in close binaries exhibit radii that are generally larger than those around single stars. Specifically, for planets in the range $0.6-1.3$ $R_{jup}$, the radii of close-binary planets are slightly smaller than those of planets in wide binaries. 
On the other hand, in the regime of $1.3-1.9~R_{Jup}$, close binaries host more inflated planets compared to both wide binaries and single-star systems.

\begin{figure}[htp]
    \centering
    \includegraphics[width=\linewidth]{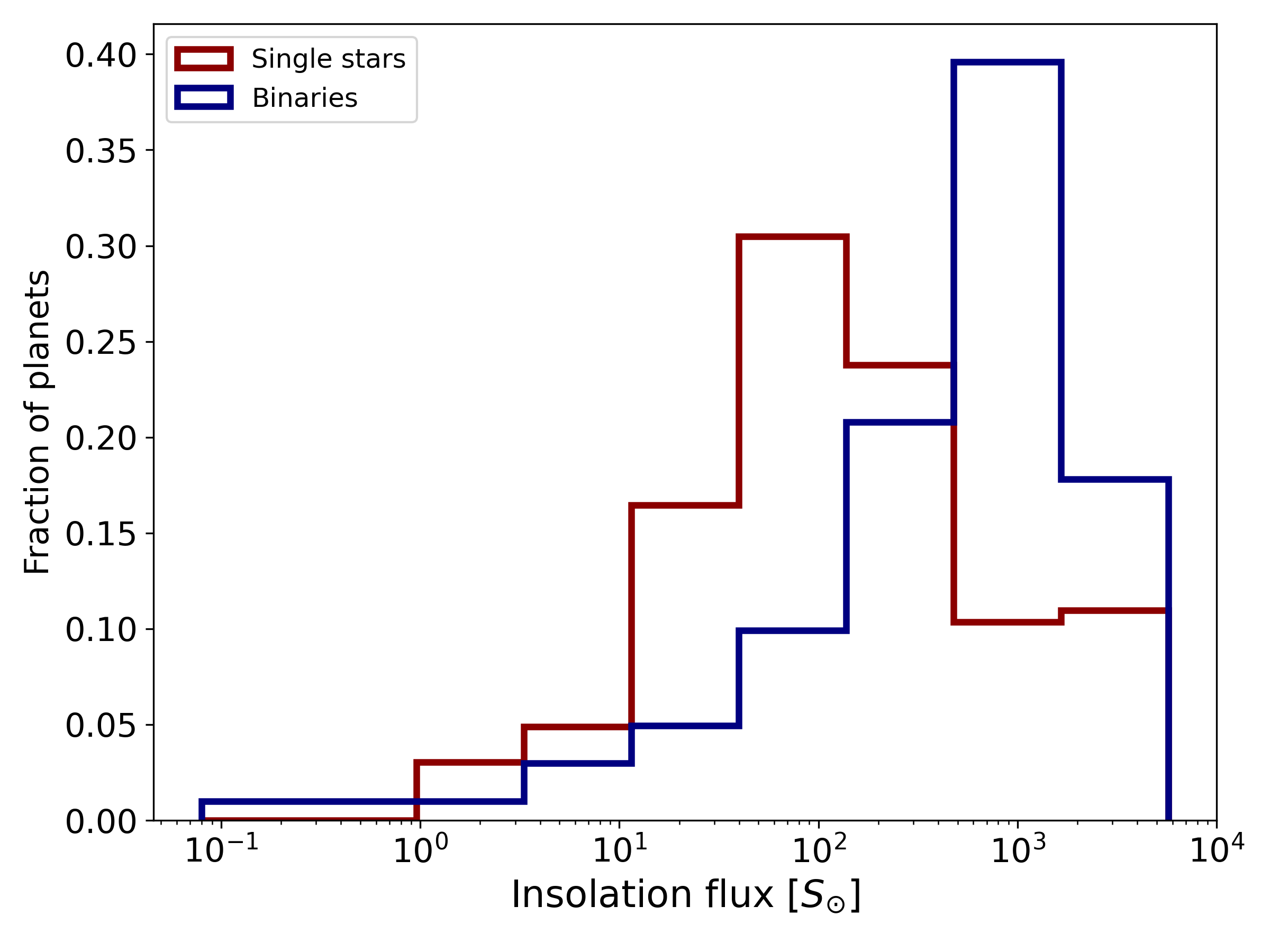}
    \caption{Insolation flux distribution for giant planets in binary systems (blue) and single-star systems (red).}
    \label{fig:ins_flux_giants}
\end{figure}

Figure \ref{fig:ins_flux_giants} shows the distribution of insolation flux for giant planets, comparing binary systems to single-star systems. Both populations span the same range of insolation; however, giant planets in binary systems seem to be skewed towards higher insolation fluxes. Consistently, the median semi-major axis of planets in binary systems ($0.045 \pm 0.002 $ AU) is smaller than that of planets around single stars ($0.050 \pm 0.001$ AU), a difference significant at the $2.7\sigma$ level (Mann-Whitney U test, p =0.028), 
consistent with enhanced inward migration of planets in binaries that are more inflated as they orbit closer to their host star and are more irradiated. This trend suggests that stellar companions may indirectly influence the radius inflation of giant planets. Rather than directly heating the planet, the companion can perturb the planetary orbit, driving high eccentricity or migration that brings the planet closer to its host star. The resulting increase in stellar irradiation and, potentially, tidal dissipation during circularisation can then contribute to the observed radius inflation of close-in giant planets \citep{Thorngren2024,2026ApJ..1001...35S}.

\begin{figure}[htp]
\centering
\includegraphics[width=\linewidth]{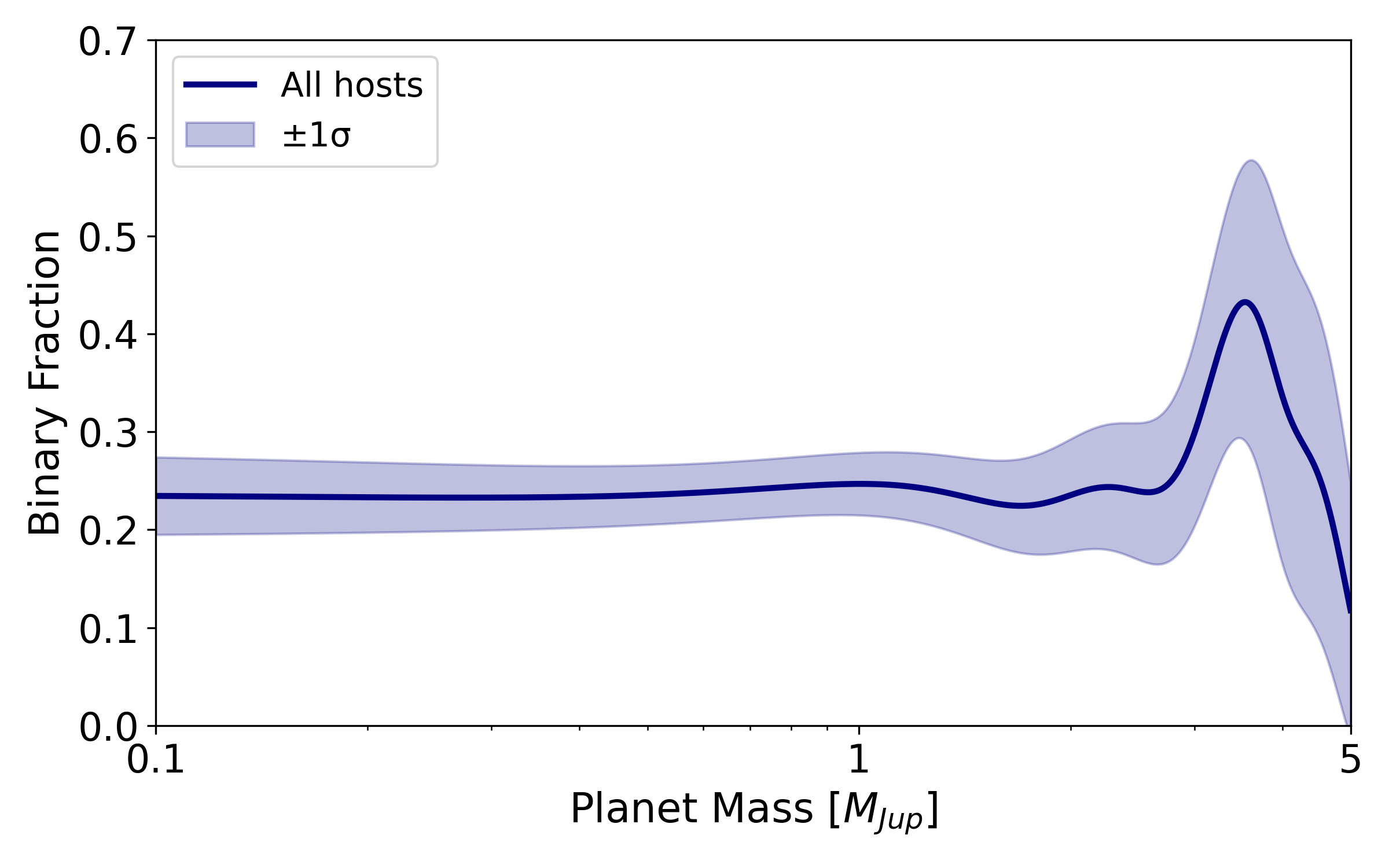}
    \caption{\small Fraction of planets at short periods (< 10 days) in binary systems as a function of planetary mass [$M_{Jup}$]. The shaded regions show the 1$\sigma$ binomial uncertainties on the weighted fractions.
          }
        \label{fig:Binaries_planet_mass}
\end{figure}

We examined the distribution of the binary fraction with respect to planetary mass, as shown in Fig. \ref{fig:Binaries_planet_mass}, for planets with orbital periods shorter than 10 days, which is the most populated region of this parameter space. We notice that the binary fraction is overall constant for masses $<~3M_{Jup}$, but the distribution exhibits a peak between 3 to 5 $M_{Jup}$. However, we find that the majority of close-in giant planets in binaries are associated with companions at wide projected separations ($\geq~2000 ~AU$, accounting for $51.87 \pm ~3.28~\%$ of systems). Aligned with previous studies \citep{2004A&A...417..353E,2002ApJ...568L.113Z,2007A&A...462..345D,2021FrASS...8...16F}, our results suggest that the most massive short-period planets tend to be found in binary systems. This might indicate that the Kozai-Lidov mechanisms play a role, but the cause remains unclear and deserves further investigation. It also suggests that the elevated binary fraction observed among hot Jupiters is not necessarily driven by tight stellar companions alone. Instead, it is consistent with recent dynamical studies showing that wide binaries, particularly when perturbed by the Galactic tidal field, can efficiently drive high-eccentricity migration over gigayear timescales, significantly contributing to the observed hot-Jupiter population \citep{2025arXiv251213773G}.\\

\subsection{Small and intermediate planets}
\label{sec: 4.3}
\begin{figure}[htp]
    \centering
    \includegraphics[width=9cm]{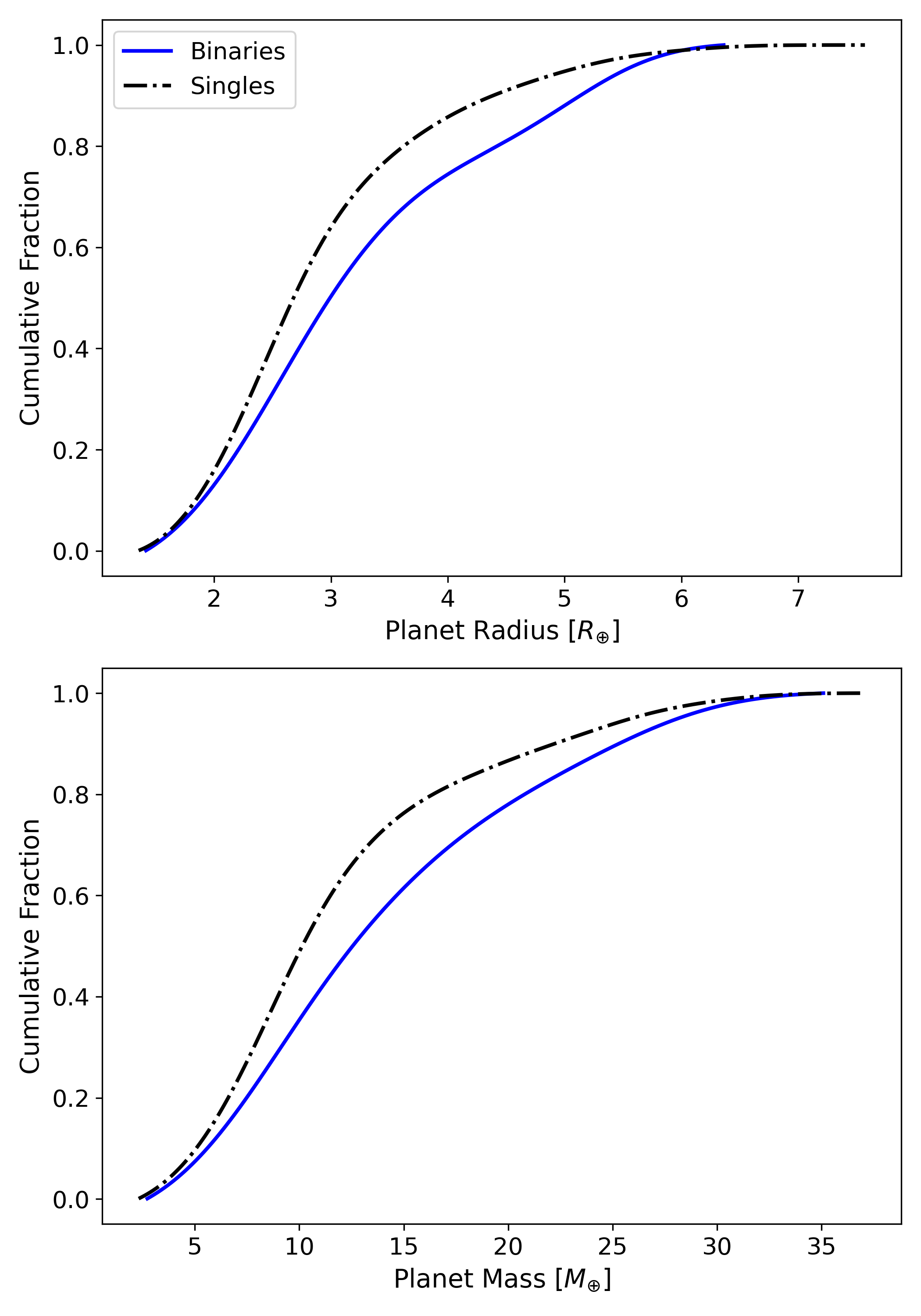}
    \caption{\small Comparison of the cumulative distribution functions of radius (top panel) and mass (bottom panel) for small-to-intermediate-sized planets in binaries (blue lines) and single stars (dash-dotted black lines). }
    \label{fig:cdf_radii_masses_region2}
\end{figure}

We extended our analysis to small and intermediate planets to investigate whether stellar binarity influences their distribution in mass and radius (Fig. \ref{fig:cdf_radii_masses_region2}). This part of the parameter space is, however, strongly affected by observational biases, particularly for planets in binary systems. The binary subsample in this regime is limited to only 30 planets compared to 170 planets orbiting single stars, corresponding to $13.78\pm 0.29~\%$ of the total sample. Super-Earths ($< 0.15 ~R_{Jup}$, or $\lesssim 1.6 ~R_{\oplus}$) are particularly underrepresented, with only three planets in binary systems. Including them in our statistics would therefore disproportionately weight small-number statistics. Excluding Super-Earths increases the binary fraction to  $17.2~\pm 3.0~\%$, but this correction does not alleviate the underlying selection effects affecting this regime. Despite these limitations, the results suggest possible differences between the two populations. In particular, planets hosted by binary systems tend to be both larger and more massive than those orbiting single stars, as illustrated in Fig. \ref{fig:cdf_radii_masses_region2}. However, recent studies show that small planets hosted in binary systems tend to be less massive than their single-star system counterparts, as the presence of a stellar companion would truncate the protoplanetary disk limiting available material to form more massive planets \citep{2026A&A...708A..38N,2026A&A...708A..37V}. As discussed in Sect. \ref{Sec. 3.2}, this region of the parameter space is subject to significant observational biases towards large and massive planets, as small planets are intrinsically more difficult to detect and characterise in binary systems. \\

\begin{figure*}[htbp]
\sidecaption
    \includegraphics[width=12cm]{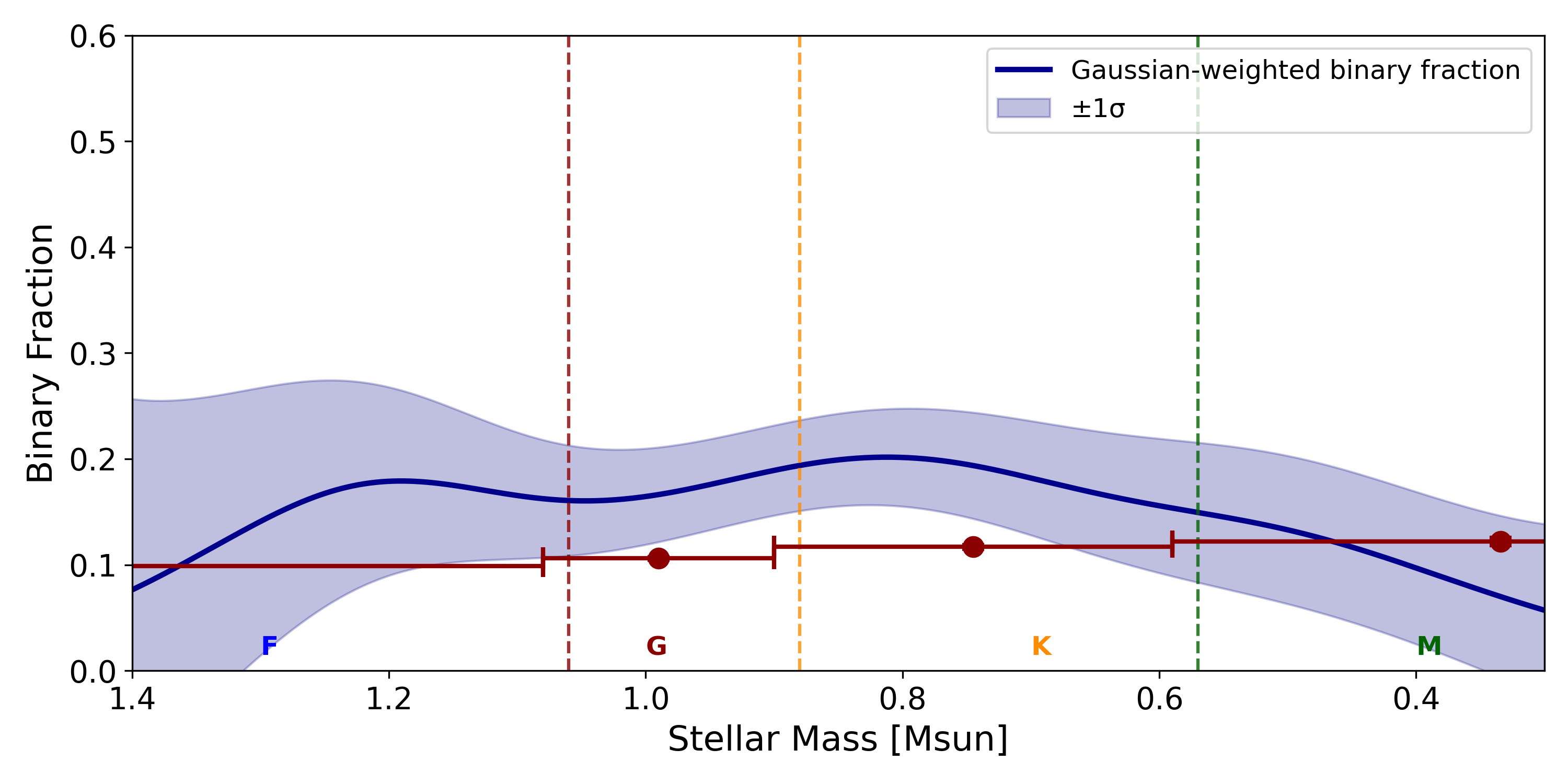}
    \caption{\small Fraction of intermediate-sized planets in binary systems as a function of stellar mass [$M_{\odot}$]. The shaded region shows the 1$\sigma$ binomial uncertainty on the weighted fraction. The vertical dashed lines mark boundaries between the F, G, K, and M spectral types, as given in \citet{2013ApJS..208....9P}. The red dots represent the known multiplicity fraction of field stars, as computed from Gaia DR3.}
    \label{fig:bin_frac_small}
\end{figure*}

Figure \ref{fig:bin_frac_small} shows the variation of the fraction of planets in binaries with respect to the total planets through stellar masses, following the same methodology as for giant planets. Once again, the fraction of field binaries in the literature is overplotted. We notice that the overall binary fraction for small planets is somewhat constant, with no statistically significant dependence on spectral type within uncertainties, with the fraction slightly decreasing for M-dwarfs, in contrast to the giant planets regime, reflecting the scarcity of detected planets in this regime for both single and binary systems. 
This apparent excess of larger and more massive planets in binary systems is most likely driven by observational biases. 

Finally, we find that the region corresponding to intermediate planetary radii $> 4~R_{\oplus}$ is sparsely populated, both for planets in binaries and for those around single stars. While the number of objects in this region in insufficient for a quantitative analysis, the observed paucity is apparent in both populations and is consistent with the well-known dearth of Neptunian planets at short orbital periods ($< 3 ~days$) often referred to as Neptunian desert \citep{2016A&A...589A..75M}, where atmospheric escape and dynamical processes are thought to efficiently erode planetary envelopes, leading to the formation of smaller-radius planets. At longer periods, the more moderately populated regions associated with the Neptunian ridge and savanna \citep{2024A&A...689A.250C,2023A&A...669A..63B} are also visible in our present sample, although it does not allow us to assess whether binarity plays a role in shaping these structures.

\subsection{Stellar properties and complementary trends}

\begin{figure}[htbp]
    \centering
    \includegraphics[width=8cm]{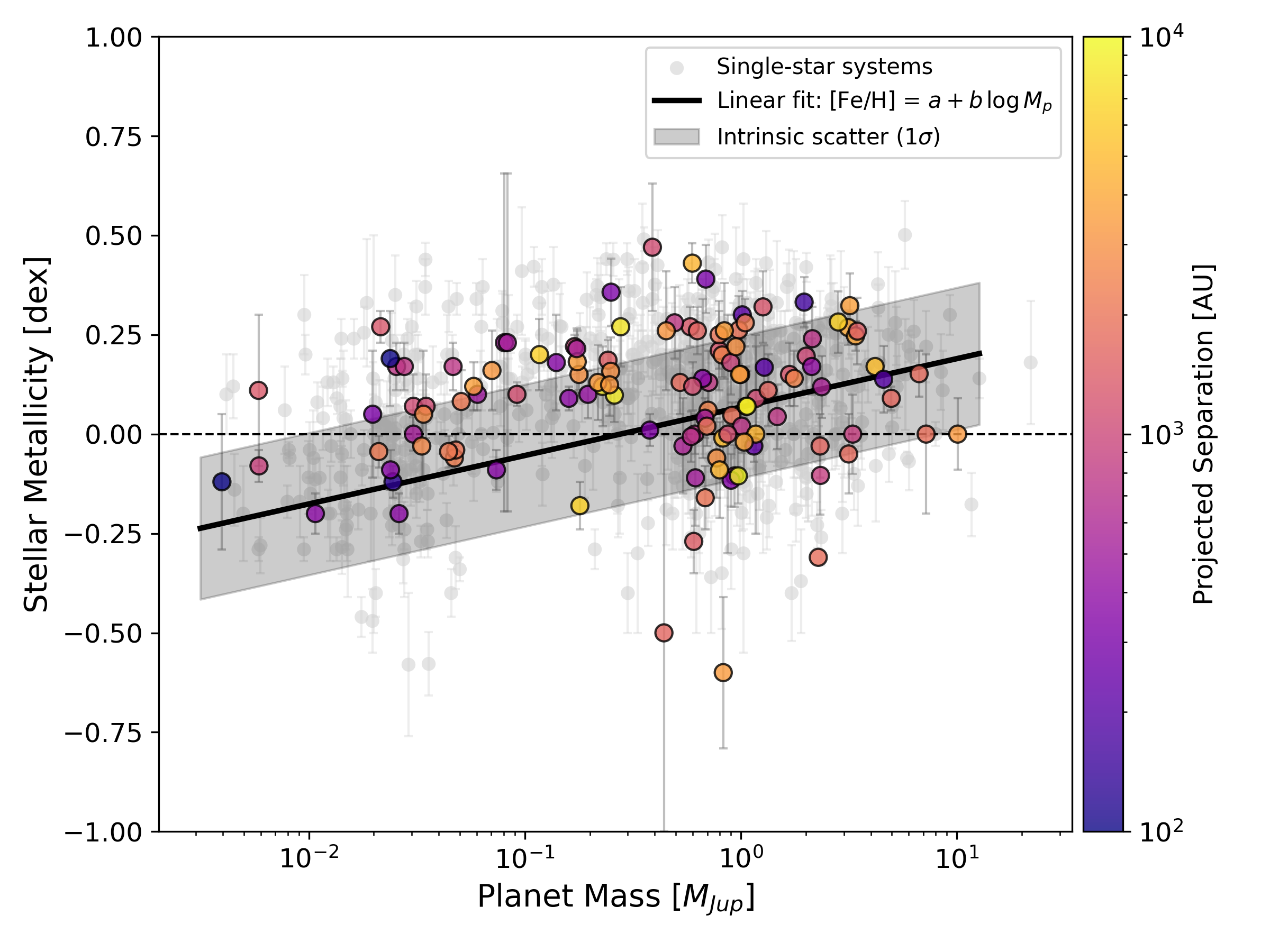}
    \caption{\small Distribution of host-star metallicities [dex] available in Gaia DR3 as a function of planet mass [$M_{Jup}$]. The binary systems are colour-coded by projected separation, while the single-star systems are colour-coded in grey. The solid line shows the best-fit linear relation between stellar metallicity and planet mass, and the shaded region indicates the $1\sigma$ intrinsic scatter around the fit. The horizontal line marks the null metallicity value. }
    \label{fig:Fe_H_hosts_Gaia}
\end{figure}

We aim to investigate a correlation between planetary and stellar characteristics for the S-type planets in our sample. 
We first examined the distribution of host-star metallicities as a function of planetary mass for all the available values on Gaia DR3, as shown in Fig. \ref{fig:Fe_H_hosts_Gaia}. Single-star systems are overplotted for comparison, while the linear regression was performed only for S-type planets.  To quantify this relationship, we modelled the correlation between host-star metallicity and planetary mass with a linear regression model in logarithmic planet mass, motivated by predictions from core-accretion theory. The shaded region indicates the intrinsic scatter of the data around the best-fit relation ($1\sigma$).

\begin{figure}[htbp]
    \centering
    \includegraphics[width=8cm]{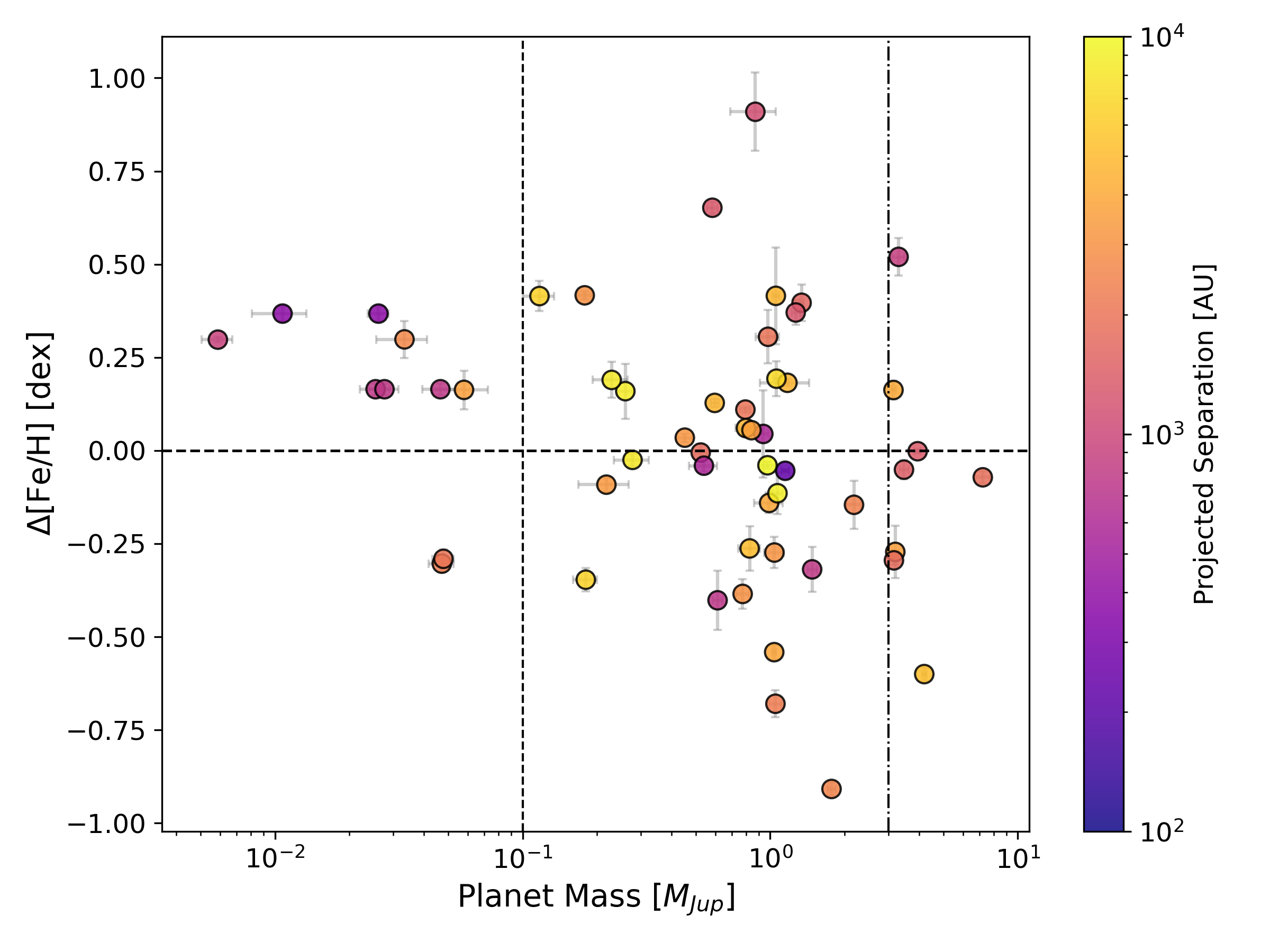}
    \caption{\small Difference in stellar metallicity between the two components of binaries with respect to planetary mass for the 59 available values in the catalogue. The horizontal line represents equal metallicity binaries. The two vertical lines delimit small to intermediate planets < $0.1 ~M_{Jup}$, giant planets $>0.1 ~M_{Jup}$, and Jupiter-like planets $>3~M_{Jup}$.}
    \label{fig:Delta_FeH}
\end{figure}

The data reveal a clear positive trend, with host-star metallicity increasing with planetary mass. This trend is largely driven by the population of giant planets, which dominate the sample at higher masses and cluster tightly around the best-fit relation. In contrast, most low-mass planets lie above the fit line, suggesting either that the regression is biased towards the giant-planet regime, where the statistics are stronger, or that different mass regimes may follow distinct metallicity dependencies. Disentangling these possibilities would require a larger and more uniformly sampled population of low-mass planets in binary systems. 
Nevertheless, the observed trend for giant planets is consistent with the core-accretion scenario, in which metal-rich protoplanetary disks more efficiently form massive solid cores, enabling subsequent rapid gas accretion and the formation of giant planets \citep{2004A&A...426L..19S,2014MNRAS.443..393G}. 
We further investigated the correlation between the planetary mass and the difference in metallicity between the stellar hosts and their companions for the 52 available values in the catalogue, as shown in Fig. \ref{fig:Delta_FeH}.
We first notice that the average metallicity difference is $\Delta[Fe/H] = 0.019 \pm 0.339 ~dex$. This indicates that for binary systems the metallicities of the two components are usually consistent since they are supposed to share the same chemical composition and be formed in the same molecular cloud. However, we notice a tentative trend in which the metallicity contrast between the planet host and its companion appears to decrease with increasing planetary mass. Given the limited size of the sample, this trend may be driven by selection effects rather than reflecting a genuine physical correlation.

For low-mass planets ($M_p<0.1~M_{Jup}$), eight out of ten systems host the planet around the more metal-rich component of the binary. This subsample includes three planets of the same system, TOI-1246, and all of these systems are composed of main-sequence stars. In several cases, such as Kepler-538 and HD 23472, the stellar companions are likely late M-dwarfs or early L-dwarfs, for which metallicity determinations are particularly uncertain. 
In contrast, most of the very massive planets ($M_p>3~M_{Jup}$) preferentially orbit the metal-poorest component of the binary, with six out of eight systems exhibiting ($\Delta[Fe/H]<0$). These planets are associated with post-main-sequence companions as confirmed by their position on the Hertzprung-Russel (HR) diagram (Fig. \ref{fig:HRD_giants_metallicities}) on the giants or sub-giant branch. This group includes the systems HAT-P-16, HD 80606, TIC 46432937, TOI-1516, WASP-14, and WASP-173 A. These two sub-populations are mainly responsible for the trend we observe; therefore, we cannot conclude robustly any correlation between the metallicity contrast of binary components and the mass of the planets they host.
A few multi-stellar systems are reported in literature to host giant planets around the apparently metal-poor stellar component. Several mechanisms have been proposed to explain such metallicity differences, all of which are plausible within current theoretical and observational frameworks (e.g. \citealt{2024MNRAS.52710016F,2016ApJ...819...19T,2021AJ....162..291J}). This highlights an important caveat: the observed metallicity contrast in these cases may not reflect the primordial chemical composition of the system but instead arise from evolutionary and formation processes. As stars evolve off the main sequence, internal mixing mechanisms such as dredge-up events can modify their surface abundances. In wide binary systems, it is therefore possible that the planet-hosting star preserves a surface metallicity closer to its initial value, while the companion becomes chemically enriched over time.
An alternative scenario involves planet engulfment by the companion star during its evolution, which could lead to an apparent metal enhancement of its outer layers. Finally, we note that systematic uncertainties in metallicity determinations for evolved stars may also contribute to the observed trend, as spectral line broadening and blending increase measurement uncertainties relative to main-sequence stars.

\FloatBarrier
\section{Discussion and conclusion}

In this work, we presented an updated and curated catalogue of transiting S-type exoplanets by cross-matching the \textit{PlanetS} database with the \citetalias{2021MNRAS.506.2269E} catalogue of wide binaries in Gaia DR3. Beyond simply increasing sample size, our approach prioritised a homogenised classification and astrophysical reliability by restricting the analysis to transiting planets with robustly derived parameters and identification of gravitationally bound stellar companions using a single, consistent Gaia-based methodology.
Using this framework, we measure a binary fraction of $19.4~\%$ when comparing transiting planets in binaries to a control sample of single-star hosts, and $15.5~\%$ when comparing to the full sample of single stars. These values are broadly consistent with previous estimates in the literature, which typically report binary fractions in the range of $15-25\%$ (e.g. \citealt{2021FrASS...8...16F,2021FrASS...8...14M,2024MNRAS.527.3183M,2023AN....34430055M,2025MNRAS.541.1419E}). However, the significance of this agreement lies not in simple confirmation but in the fact that it is obtained from a more homogeneous and controlled sample than previously available. 
Most earlier demographic studies of planets in binary systems such as \citet{2025A&A...700A.106T} rely on heterogeneous samples that combine planets detected via multiple techniques (radial velocity, transit, direct imaging, and high-resolution imaging surveys), often within limited volumes ($\sim$20-25 pc). While powerful, such compilations inherently mix detection biases, selection functions, and parameter uncertainties that are difficult to model consistently. As a result, the inferred binary fractions and demographic trends are sensitive to survey incompleteness and method-dependent biases, particularly at small planetary radii and close visual binary separations.
In contrast, our analysis isolates transiting systems enabling a statistically cleaner comparison, drawn from a single, internally consistent catalogue; it applies uniform criteria to both planet and host properties and a binary identification performed using a uniform Gaia DR3-based methodology, enabling a statistically cleaner comparison between single and binary host populations.

We emphasise that our identification of binary systems is intentionally conservative. By restricting the sample to confidently resolved, gravitationally bound companions in Gaia DR3, we necessarily obtain a lower limit on the true binary fraction in our sample. This approach ensures that the binary sample is robust and minimises contamination by chance alignments or spurious associations. We note, however, that our sample remains subject to detection and selection biases that preferentially favour large and massive planets. Nevertheless, in the giant-planet regime (as defined in Fig. \ref{MRfig}), these biases are significantly reduced compared to smaller planets, making our statistical inferences in this domain more reliable.

Focusing on the regime of massive planets with reliable detection statistics, we obtain a binary rate of $\sim 22.08~\%$ ($17.68~\%$ if compared to the full sample of single-star systems), and we observe noticeable differences as a function of host spectral type and binary separation. In particular, we find that, when hosted by M-dwarfs, giant planets tend to be preferentially in close binary systems with a projected separation < 1000 AU. Although the number of such systems remains limited, this trend is noteworthy given that giant planet formation around isolated low-mass stars is expected to be inefficient and suggests that stellar multiplicity may play a role in shaping the occurrence or evolution of massive planets around M-dwarfs.
We caution, however, that the projected separation is not equivalent to the true orbital separation of the binary. Previous work \citep[e.g.][]{1935PASP...47...15K,1935PASP...47..121K,1991A&A...248..485D}
has shown that for wide binaries the random orbital orientations and moderate eccentricities, the semi-major axis $a_{bin}$ can be statistically approximated from the projected separation $\rho_{bin}$ as $\langle \rho \rangle = f\,a$, where \citep{1935PASP...47...15K,1935PASP...47..121K} empirically defined $f = 0.776$ from a large sample of different binary configurations. We note that for the majority of the systems in our sample the full 3D binary orbits are not yet resolved. Future Gaia releases, in particular DR4 and DR5, will deliver improved astrometric solutions and orbital constraints for wide binaries, enabling the reconstruction of true orbital architectures and enabling a more physically grounded interpretation of binary-driven effects on planet formation.

Among the planets in our sample hosted in stellar binary systems, only five orbit the secondary component of the binary (less massive component), corresponding to only $3.7~\%$ of the planets in binaries. All of these systems have angular separations ranging from 11" to 40", suggesting that this low fraction is most likely driven by observational biases rather than astrophysical suppression, as flux contamination from the primary is negligible at such wide separations.

Beyond the results presented here, several forthcoming facilities will substantially improve our ability to characterise planets in binary systems. In particular, on-going efforts with the
\textit{Near-InfraRed Planet Searcher (NIRPS)} \citep{2025A&A...700A..10B} as part of its Guaranteed Time of Observation (GTO) aim to extend radial-velocity follow-ups of transiting planets around low-mass stars in binary systems. Its configuration enhances angular resolution and mitigates flux contamination from close stellar companions, making \textit{NIRPS} particularly well suited to probing the close-binary regime (< 2 arcsec) and delivering accurate mass measurements for transiting planets around M-dwarfs. 

In parallel, upcoming Gaia DR4 and DR5 releases will provide improved astrometric solutions and orbital constraints, allowing us to probe tighter angular separations than currently possible as well as identify additional new binaries that can yet not be resolved. Meanwhile, the PLATO mission will deliver a large, homogeneous sample of transiting planets around bright stars, offering an unprecedented opportunity to expand demographic studies of S-type planets. Additionally, it will provide asteroseismology data that would be helpful for better characterising the stellar companions. Furthermore, the 3D orbital architecture of binary systems hosting planets, in particular the mutual inclination between planetary and binary orbits, represents an additional dimension of this problem that has not been explored in this work. Recent studies suggest that planet-binary systems tend to be nearly aligned, with a possible misaligned population emerging at larger binary periastron distances \citep{2023AJ....166..166L,2026AJ....171...77Z}. Future Gaia DR4 and DR5 astrometric solutions combined with high-resolution imaging will allow us to extend such investigations to the broader sample presented here. In anticipation of these forthcoming datasets, the catalogue introduced here serves as a robust and reliable community resource, designed to facilitate future investigations of exoplanet demographics with stellar multiplicity placed at the core of the analysis.

\begin{acknowledgements}
This work has been carried out within
the framework of the NCCR PlanetS supported by the Swiss National Science
Foundation under grants 51NF40\_182901 and 51NF4\_205606. JV and AN acknowledge support from the Swiss National Science Foundation (SNSF) under grant PZ00P2\_208945.
We thank Dr. Kareem El-Badry for his helpful correspondence and valuable advice regarding our binaries identification procedure and the computation of the binary fraction among Gaia DR3 field stars.\\

\end{acknowledgements}

 \bibliographystyle{aa}
 \bibliography{Bibliography}

\begin{appendix}
    \section{Reassessment of triple systems in the \textit{PlanetS} catalogue}
    \label{Appendix A}

The original version of the \textit{PlanetS} catalogue included 15 systems classified as triple stellar systems. We reassessed each of these systems using Gaia DR3 astrometry, and available literature information to ensure a consistent and physically motivated adopted classification of stellar multiplicity in the scope of the present work.

As shown in Table \ref{tab:triple_systems}, we retained only one system, LTT-1445, as a confirmed triple in our catalogue, The remaining 14 systems were reassessed within the scope of this work using a conservative, Gaia-based classification scheme. Systems listed in the \citetalias{2021MNRAS.506.2269E} catalogue, which does not include triple systems, were treated as binaries (Number of companions = 1). For systems not included in that catalogue, we only considered resolved pairs or triples in Gaia DR3 with consistent parallaxes and proper motions, or systems with reliable astrometric solutions in the literature. In the absence of resolved Gaia detections or astrometric confirmation for additional components, systems were treated as single stars for the purposes of this analysis. This approach prioritises robustness and homogeneity, at the cost of potentially underestimating higher-order multiplicity. 

  \begin{table}[h!]
    \footnotesize
    \centering
    \caption{Previously identified triple systems and their updated classification in the PlanetS catalogue.}
    \begin{tabular}{llp{2.5cm}}
    \hline
    \textbf{System ID} & \textbf{Source} & \textbf{Number of companions adopted} \\
    \hline
    BD-14 3065 A & \text{\citet{2024A&A...688A.120S}} & 0 \\
    HAT-P-16 & \text{\citet{2016MNRAS.459..789T}} & 0 \\
    HAT-P-35 & \text{\citet{2012AJ....144...19B}} & 0 \\
    HAT-P-8 & \text{\citet{2017A&A...602A.107B}} & 0 \\
    K2-27 & \text{\citet{2017AJ....153..142P}} & 0\\
    K2-290 & \text{\citet{2019MNRAS.484.3522H}} & 1 \\
    KELT-4 A & \text{\citet{2016AJ....151...45E}} & 1 \\ 
    KOI-13 &\text{\citet{2014ApJ...788...92S}} & 0 \\
    LTT 1445 A & \text{\citet{2023A&A...673A..69L}} & 2  \\
    TOI-2152 & \text{\citet{2023MNRAS.521.2765R}} & 1 \\
    WASP-11 & \text{\citet{2009ApJ...696.1950B}} & 0 \\
    WASP-12 & \text{\citet{2017AJ....153...78C}} & 0  \\
    WASP-14 & \text{\citet{2017A&A...602A.107B}} & 1 \\
    WASP-24 & \text{\citet{2017A&A...602A.107B}} & 1 \\
    WASP-3 & \text{\citet{2010MNRAS.405.1867S}} & 0 \\
    \hline
    \end{tabular}
    \label{tab:triple_systems}
    \end{table}

\newpage

    \section{RUWE distributions of planet hosts and stellar companions}
    \label{Appendix ruwe}

    As discussed in Sect. \ref{Sec. 3.2}, the RUWE provided in Gaia DR3 offers a complementary diagnostic for identifying potentially unresolved stellar companions. It quantifies the quality of a single-star astrometric fit and is described in details in \cite{2021A&A...649A...2L}. A well-behaved source with a good single-star astrometric solution is expected to correspond to RUWE close to $1.0$, while elevated values suggest that the single-star model is a poor fit to the astrometric data, possibly due to the presence of an unresolved companion \citep{2018A&A...616A...2L,2021A&A...649A...2L}.

    The choice of a specific RUWE threshold is not straightforward, as different studies have adopted different values depending on the Gaia data release and the characteristics of the observed sample. \citet{2018A&A...616A...2L} originally proposed a threshold of 1.4 within Gaia DR2 to identify reliable astrometric solutions, while \cite{2022MNRAS.513.2437P} recommended a more conservative threshold of 1.25 for identifying potential binaries in Gaia EDR3, based on the local renormalised unit weight error (LUWE). In light of this, we examined our sample using both thresholds to assess the sensitivity of our results to this choice.

    \begin{figure}
        \centering
    \includegraphics[width=\linewidth]{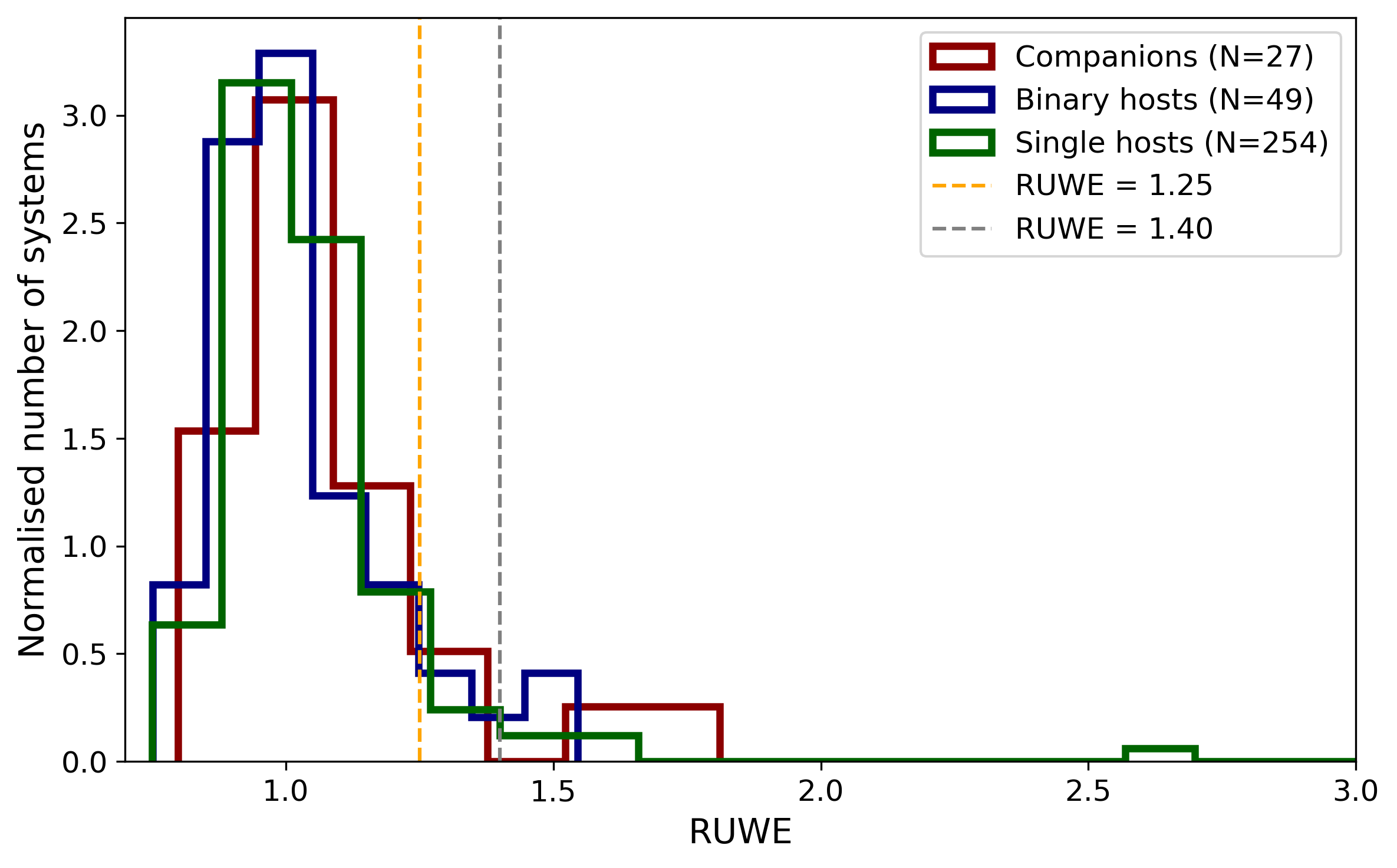}
    \caption{\small Distributions of RUWE values available in Gaia DR3 for stellar companions (red), binary hosts (blue), and single-star hosts (green). The dashed lines represent the thresholds adopted in the literature for the elevated RUWE values at 1.25 (yellow) and 1.4 (grey) \citep{2021A&A...649A...2L,2022MNRAS.513.2437P}.}
    \label{fig:ruwe_distributions}
    \end{figure}

    Figure \ref{fig:ruwe_distributions} shows the RUWE distributions from Gaia DR3 for the populations of binary stellar hosts, their identified companions, and the single-star hosts. We note that RUWE values were available for only 49 out of 123 binary hosts, 31 out of 123 stellar companions, and 262 out of 725 single-star hosts, reflecting the incompleteness of the Gaia DR3 astrometric solution for a fraction of our sample.

    The three populations show overall well-behaved astrometric solutions, with median RUWE values of 1.004, 1.036, and 1.012 for the binary hosts, their companions, and the single-star hosts respectively. Only four stellar companions and five binary hosts have a RUWE > 1.25, while only 26 single systems have a RUWE > 1.25.
    Overall, the low fraction of elevated RUWE values across all three populations suggests that the vast majority of our sample has reliable Gaia DR3 astrometric solutions.
    The small fraction of single-star hosts with elevated RUWE values suggests that some systems classified as single in our sample may in fact host unresolved stellar companions. As a consequence, the binary fractions reported in this study should be interpreted as lower limits of the true binary fraction in our sample.
    
    \section{Hertzsprung-Russell diagram for evolved companions}
    \label{Appendix HRD}

    \begin{figure}[h!]
    \centering
    \includegraphics[width=0.8\linewidth]{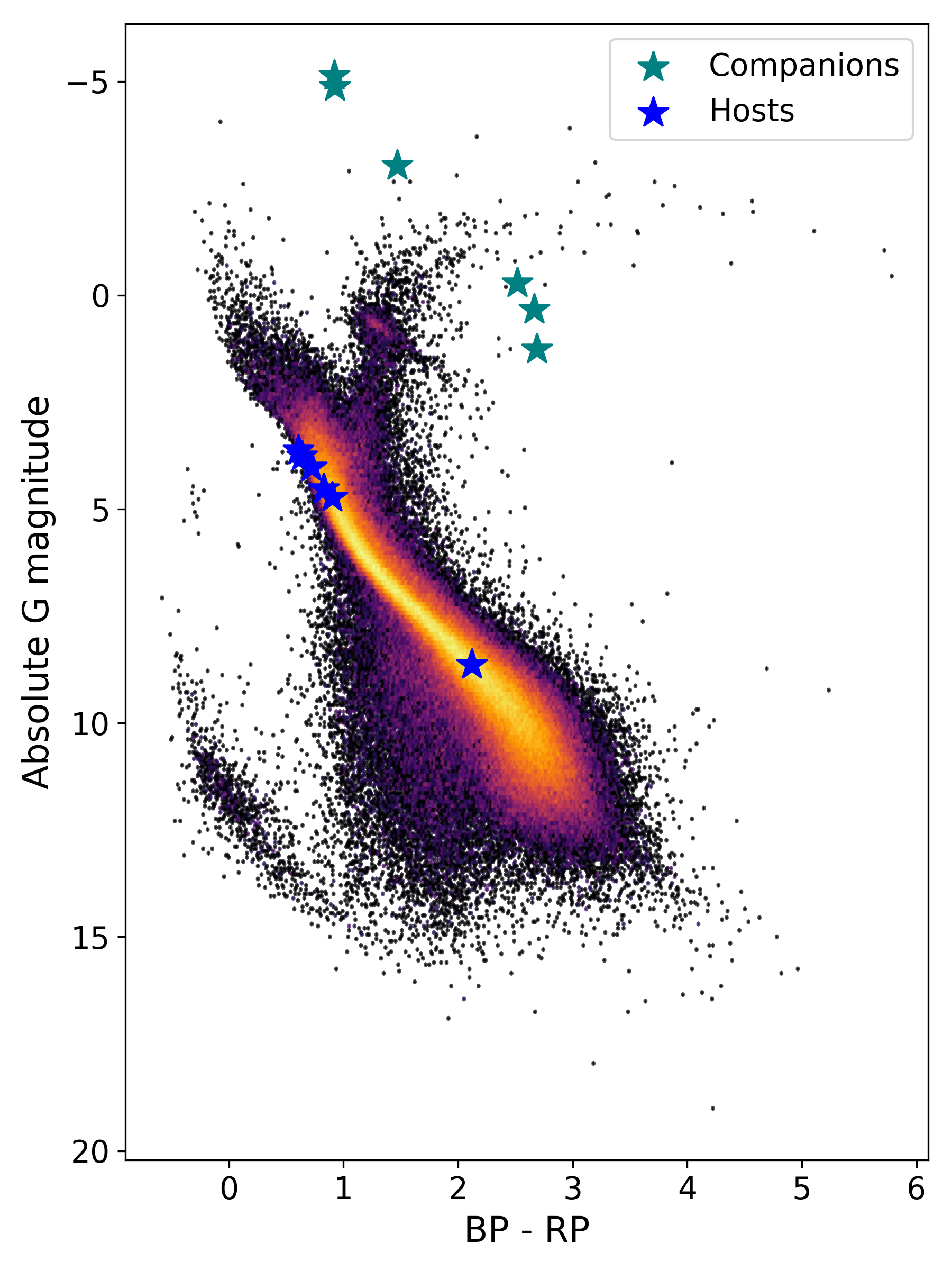}
    \caption{\small Hertzsprung–Russell diagram constructed from a representative subset of Gaia DR3 sources. The absolute magnitude in the G band is plotted as a function of the BP–RP colour and colour-coded using point densities. The star-shaped markers represent the host stars of giant massive planets ($M_{p} >3~M_{Jup}$), which are the metal-poorest components of the binaries (blue), and their stellar companions (teal).}
    \label{fig:HRD_giants_metallicities}
    \end{figure}

\end{appendix}

\end{document}